# How Explainable AI Affects Human Performance: A Systematic Review of the Behavioural Consequences of Saliency Maps

Romy Müller

*Faculty of Psychology, Chair of Engineering Psychology and Applied Cognitive Research, TUD Dresden University of Technology, Dresden, Germany*

Corresponding author:

Romy Müller

Chair of Engineering Psychology and Applied Cognitive Research

TUD Dresden University of Technology

Helmholtzstraße 10, 01069 Dresden, Germany

Email: romy.mueller@tu-dresden.de

Phone: +49 351 46335330

ORCID: 0000-0003-4750-7952

## Abstract

Saliency maps can explain how deep neural networks classify images. But are they actually useful for humans? The present systematic review of 68 user studies found that while saliency maps can enhance human performance, null effects or even costs are quite common. To investigate what modulates these effects, the empirical outcomes were organised along several factors related to the human tasks, AI performance, XAI methods, images to be classified, human participants and comparison conditions. In image-focused tasks, benefits were less common than in AI-focused tasks, but the effects depended on the specific cognitive requirements. Moreover, benefits were usually restricted to incorrect AI predictions in AI-focused tasks but to correct ones in image-focused tasks. XAI-related factors had surprisingly little impact. The evidence was limited for image- and human-related factors and the effects were highly dependent on the comparison conditions. These findings may support the design of future user studies.

## 1 Introduction

Explainable artificial intelligence (XAI) holds the promise of making deep neural networks more transparent to humans. In the context of image classification, the most common XAI approaches are attribution methods. They try to infer which image areas the AI has used for classification and then highlight these relevant areas via saliency maps (see Figure 1 for examples). Initially, great hopes have been placed on such saliency maps. Understanding which features determine the AI model's classification could give an immense boost to human-AI collaboration. This is because the powers of complex image classifiers could be harnessed while still keeping humans in the loop, thus enabling a higher overall system performance. But are saliency maps really useful for humans?

Answering this question is complicated by the fact that the usefulness of saliency maps cannot be inferred from automated metrics of XAI quality. Their results typically diverge from those obtained in user studies that require humans to actually work with saliency maps. This is the case both for metrics that assess an XAI method's fidelity to the AI model (Biessmann & Refiano, 2019; Colin et al., 2022; Lu et al., 2021; Müller, Thoß, et al., 2024) and for those that assess its similarity to human-selected image areas (Kim et al., 2022; Nguyen et al., 2021). Accordingly, user studies seem inevitable in the evaluation of XAI, if the aim is to generate explanations that can support human decision making. However, recent findings from user studies have severely dampened the initial optimism about the usefulness of saliency maps. While they were often found to increase self-reported human trust in the AI, they did not actually improve human-AI collaboration (e.g., Kim et al., 2022). They were appealing but not useful when it comes to objective performance. In fact, they may even lead to characteristic biases due to misinterpretations of feature relevance (Balayn et al., 2022; Kim et al., 2018) or an uncritical agreement with incorrect AI predictions (Nguyen et al., 2021; Stock & Cisse, 2018).

Obviously, saliency maps do not always have these adverse effects. Instead, a huge variety of empirical outcomes can be observed across user studies. Sometimes saliency maps had no effects on the accuracy of human performance (Nguyen et al., 2021), while at other times they were helpful for some images but harmful for others (Sayres et al., 2019). Similarly, saliency maps sometimes were helpful only when the AI predictions were correct (Kim et al., 2022) and sometimes only when they were incorrect (Alipour et al., 2020; Yang et al., 2022). Such diverging outcomes are abundant in the literature. When trying to make sense of them, it should be considered that different XAI methods were used in different domains that rely on different image data, aiming to support different human tasks performed by different human beings. Given this immense variability, it is not surprising that the outcomes diverge. Is it still possible to derive an integrated picture of the conditions under which saliency maps enhance human performance? The present review set out to meet this challenge.

Reviews of XAI user studies have become increasingly popular in recent years (Bertrand et al., 2022; Fok & Weld, 2023; Haque et al., 2023; Kandul et al., 2023; Rong et al., 2023; Schemmer et al., 2022). However, the focus of these reviews has usually been rather broad. In an attempt to cover the entire field of XAI, they included all types of XAI methods used on all types of data. Moreover, most reviews assessed the effects of XAI on humans in a variety of dependent variables, including subjective ratings of trust or XAI quality as well as measures of understanding. Such broad reviews necessarily have to remain superficial regarding the effects of specific types of XAI on specific aspects of human-AI interaction. Accordingly, previous reviews usually did not include more than a handful of studies that investigated how saliency maps affect human performance in the context of image classification. The present review is more focused: it only considered empirical user studies that applied attribution methods to generate saliency maps for images and that report how these saliency maps affected human performance outcomes. Specifically, the article makes the following contributions:

- A systematic review of 68 empirical user studies from 52 publications, investigating the effects of saliency maps on human performance in image classification and related tasks.
- To determine how the effects of saliency maps depend on contextual modulators, the empirical outcomes were organised along relevant factors related to the human task, AI, XAI, images, humans, and the investigated comparisons.
- To facilitate subsequent use and extension of the present findings, the research materials are made available. This includes a table characterising each of the 68 empirical studies on 70 variables, hopefully enabling other research to answer their own questions based on the same data material.

To provide an overview of related work, the following section introduces previous reviews of user studies on the effects of XAI (Section 2). It is explained why it is worthwhile to sharpen the focus and how this is accomplished in the present review (Section 3). After summarising the review methods (Section 4), the results are presented (Section 5). The presentation of results starts out with a descriptive overview of the included studies concerning their XAI methods, human tasks, dependent variables and other aspects (Section 5.1). This is followed by an overview of the empirical outcomes and obtained biases (Section 5.2). The main part extracts the modulators of XAI effects and organises the empirical outcomes along these factors (Section 5.3). In the final section, the results are integrated and discussed in the light of previous literature, the limitations of the present review are made transparent and perspectives for future research are outlined (Section 6).

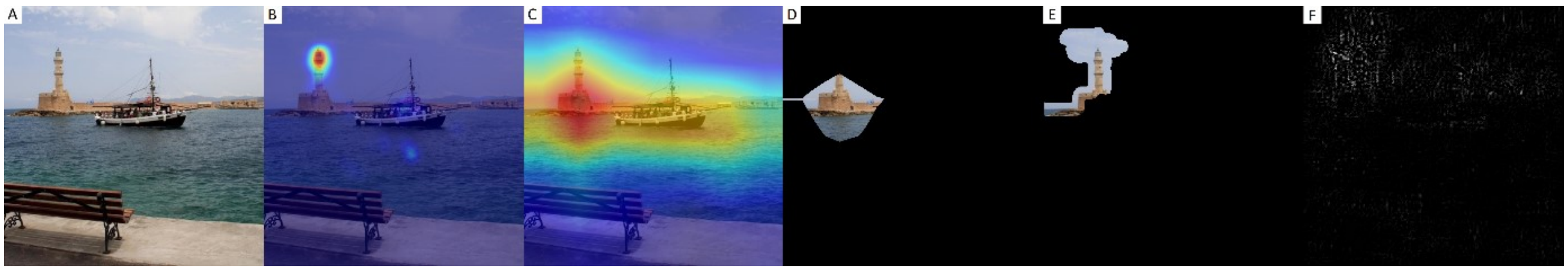

**Figure 1.** Saliency maps for the class "lighthouse" along with (A) the original image and (B) the areas considered most relevant by humans (ground truth estimated via eye tracking). (C) Heatmap generated via Grad-CAM (Selvaraju et al., 2017). The remaining images show segments reflecting the most relevant 5% of the image according to three attribution methods: (D) Grad-CAM, (E) XRAI (Kapishnikov et al., 2019) and (F) Integrated Gradients (Sundararajan et al., 2017).

## 2 Related work

Given the flood of recent XAI publications, it has become increasingly important to systematise and integrate their findings. This has resulted in an increasing popularity of literature reviews. Thus, the contributions of the present review can best be specified by contrasting it with three other types of reviews.

First, there are numerous reviews of XAI methods that also include user studies in their collections of reviewed evaluation methods (e.g., Eldrandaly et al., 2023; Mohseni et al., 2021; Vilone & Longo, 2021). However, given that the impacts of XAI on human users is just one of several issues addressed in these reviews, they can only provide a rather limited selection of the empirical outcomes.

Second, there are prescriptive reviews that do start from a user perspective but then focus on what *should* be done, instead of describing empirical findings in detail. That is, these articles provide design recommendations, either for explanations (e.g., Laato et al., 2022) or for user studies (e.g., Davis et al., 2020; Scharowski et al., 2022). For instance, Davis et al. (2020) come to the conclusion that user studies should focus on utility rather than trust. That is, they suggest assessing the performance effects of XAI in so-called "downstream tasks" instead of assessing how XAI is subjectively evaluated. While such prescriptive articles may make convincing arguments, they are usually based on theoretical considerations combined with fragmented observations of empirical outcomes. In contrast, descriptive reviews strictly base their implications for future research on a systematic account of the actual outcomes of past research. Following this tradition, the present review relies on a bottom-up, descriptive instead of prescriptive approach, empirically analysing what has actually been done in XAI user studies instead of theoretically analysing what should be done.

The third type of review is concerned with the empirical effects of XAI on human users (e.g., Bertrand et al., 2022; Fok & Weld, 2023; Haque et al., 2023; Kandul et al., 2023; Rong et al., 2023; Schemmer et al., 2022). Perhaps the most comprehensive review to date has been presented by Rong et al. (2023). They considered a wide range of XAI methods for different tasks and data types as well as a wide range of dependent variables, including measures of trust, understanding, usability and performance. A central aim of their review was to analyse how user studies are currently implemented and provide suggestions for future studies. To manage this thematic breadth, the authors only performed a coarse-grained automatic literature search of selected conferences. In consequence, the number of included publications that dealt with the performance effects of saliency maps for image classification is rather low. While the authors report mixed findings regarding the effects of XAI on performance, the small number and large variability of included studies does not allow them to explain these inconsistencies. A similarly broad review has been presented by Haque et al. (2023). However, the authors did not consider objective performance at all but focused on the effects of XAI on subjective measures like trust, transparency, understandability, usability and perceived fairness. Kandul et al. (2023) did consider performance and concluded that it is rarely enhanced by XAI. However while they included 11 publications that dealt with image classification, many of them investigated other types of XAI than saliency maps and other dependent variables than performance.

Other reviews set a more specific focus. For instance, Bertrand et al. (2022) asked how cognitive biases affect XAI-assisted decision making. They collected 53 biases mentioned in the XAI literature (including biases that affect the design and evaluation of XAI as well as those that can be mitigated or exacerbated by XAI). However, they considered a wide range of XAI approaches (e.g., rule-based or counterfactual explanations) and only included two articles about saliency maps. Finally, two reviews specifically investigated whether XAI improved human performance (Fok & Weld, 2023; Schemmer et al., 2022). Both of them found that it did not. Fok and Weld (2023) suggested that XAI only is helpful to the extent that it allows users to verify AI predictions. Schemmer et al. (2022) reported XAI benefits compared to completely unassisted human performance but none compared to AI without explanations. However, both reviews included a wide variety of XAI approaches, data types and tasks. Moreover, both only included two or three image classification studies. Thus, to date no conclusions can be drawn about the performance effects of saliency maps in the context of image classification.

## 3 Present study

Previous reviews of the impacts of XAI on human users are limited by their broad focus. They usually considered different types of data and tasks, different XAI approaches and different dependent variables to describe the effects of XAI. This thematic breadth comes with benefits in terms of generality but also costs in terms of information depth. To date, it is unclear how saliency maps affect human performance and what contextual factors modulate these effects. Closing this research gap is worthwhile for three reasons.

First, it seems reasonable to evaluate explanations based on how they affect objective human performance (Davis et al., 2020). This is because the outcomes of subjective and objective measures may diverge systematically. While XAI is often evaluated positively by its users, this does not always go along with performance benefits (Fok & Weld, 2023; Kandul et al., 2023). Similarly, the superiority of one XAI method over another may change or even reverse depending on whether it is assessed via subjective evaluations or objective performance (Buçinca et al., 2020). Accordingly, the usefulness of XAI cannot reliably be inferred from subjective ratings of trustworthiness or even usability.

Second, focusing on image classification might generate insights that differ from those obtained with other data and tasks. One reason is the presumed specificity of XAI-related biases (Bertrand et al., 2022). Many of them do not seem applicable in the context of saliency maps for image classification (e.g., biases associated with memory, understanding language or integrating probabilities). In addition to that, many image classification tasks may not share the moral or legal relevance of other tasks investigated in the XAI literature (e.g., university admissions, recidivism prediction or fraud detection). Image classification seems like a rather straightforward, objective task for which the correctness of AI predictions can be evaluated unambiguously. It may not require users to draw inferences beyond the visible data and gain a deeper understanding of the data beyond what they can already see without XAI. This may affect how users approach the task and rely on the XAI. However, it is unclear whether it will result in a higher or perhaps even lower propensity of performance benefits and a more or even less direct match between subjective and objective measures.

Third, in the context of image classification, it seems worthwhile to focus on attribution methods that generate saliency maps to explain feature relevance. Attribution methods are among the most frequently applied types of XAI. This can be illustrated by the popularity of Grad-CAM (Selvaraju et al., 2017), which is just one particular method for generating saliency maps and yet it has been cited more than 20,000 times to date. Despite their popularity, saliency maps are limited in the information they convey (Colin et al., 2022): they can only tell users *where* the AI is attending but not *what* it is doing with the information. Moreover, saliency maps have been criticised for not enabling users to verify the AI predictions, which was assumed to be a necessary condition for XAI to be useful (Fok & Weld, 2023). Thus, if saliency maps are useful at all, this is likely to be restricted to specific tasks that pose specific cognitive requirements. Presumably, reasoning about cognitive requirements is not the key competence of many XAI researchers who do not have a psychological background. Therefore, knowledge about the boundary conditions under which saliency maps can be useful should be made available to the XAI community.

In the present study, a systematic review investigated how saliency maps affect human performance. More precisely, it aimed to specify how the effects of saliency maps are modulated by a number of contextual factors. *Task-related factors* refer to the general focus of human tasks (i.e., focusing human attention either on the AI or the image) as well as the specific task assignments that users have to perform (e.g., predict the AI prediction, report the image class). Moreover, as it will turn out below, only considering these concrete task assignments is insufficient and a categorisation of tasks based on their underlying cognitive requirements is needed. The second type of modulators, *AI-related factors*, refers to aspects of AI performance such as its accuracy or the type of bug. When considering *XAI-related factors*, the present review does not intend to evaluate the superiority of any particular XAI method. Rather, it asks to what extent the selection of XAI methods affects performance, considering factors such as saliency map quality and visualisation. *Image-related factors* refer to the characteristics of images that affect the results, such as image complexity or the perceptual similarity of classes. *Human-related factors* are investigated to reveal whether saliency maps are more useful for some people than others, for instance depending on their level of expertise. Finally, *comparison-related factors* may affect the results. Saliency maps can be compared to different baselines such as AI support without explanation, other types of XAI or extensions of traditional saliency maps. Depending on this comparison, saliency maps may be more or less useful.

## 4 Methods

### 4.1 Inclusion and exclusion criteria

The present review aimed to investigate how saliency maps affect human performance. It focused on attribution methods for image classification that analyse feature relevance and visually highlight the areas affecting the predictions of deep neural networks (DNN). The main goal was to create a nuanced and representative overview of the available studies by highlighting the context-specificity of XAI effects. With this goal in mind, the inclusion and exclusion criteria were defined as listed in Table 1.

**Table 1.** Inclusion and exclusion criteria.

| Type | Inclusion criteria | Exclusion criteria |
|---|---|---|
| Type of XAI | • Attribution methods for DNN that visualise feature relevance via saliency maps<br>• Highlighting of relevant image areas (e.g., heatmaps, luminance maps, segments, outlines)<br>• Extensions of saliency map methods and combinations with other types of XAI<br>• Local explanations, model-agnostic and non-agnostic methods<br>• Real AI and XAI | • Other types of XAI (e.g., concept-based, rule-based)<br>• Only other XAI output (e.g., text, synthetic images, image examples from the dataset)<br>• Global explanations<br>• Simulated XAI, highlighting that is not based on DNN |
| Data | • Images of physical objects, computer-generated images, handwriting, visual art<br>• Photographs and other imaging technologies (e.g., X-ray, sonar) | • Other data types (e.g., textual, tabular, videos, 3D animations)<br>• Data visualisations (e.g., graphs, time series) |
| Type of evaluation | • Empirical evaluations with human participants (e.g., experiments, observations, interviews) | • Evaluations based on automated metrics<br>• Evaluation by the authors |
| AI task | • Image classification<br>• Related tasks (e.g., object detection, visual question answering) | • Other image-dependent tasks (e.g., image retrieval, image generation) |
| Human task | • AI-focused (e.g., predicting AI predictions or AI accuracy, detecting AI biases)<br>• Image-focused (human-AI collaboration, see AI task) | • Evaluating XAI visualisations<br>• Indicating preference of XAI method<br>• Self-reports (see dependent variables) |
| Comparison conditions | • AI without explanations<br>• XAI substitutes (e.g., bottom-up image saliency)<br>• Other types of XAI (e.g., concept-based)<br>• Different versions of saliency maps or extensions<br>• Other control conditions (e.g., random guessing) | • No control conditions<br>• Comparisons that do not include pure saliency maps (e.g., including only combined XAI methods) |
| Dependent variables | • Human task performance (e.g., accuracy, solution time, agreement with the AI, detection of undesired AI behaviour) | • Subjective assessment of XAI quality (e.g., trust, usability, understandability, visualisation quality, preference)<br>• Understanding, mental models |
| Reporting standards | • Meets minimum requirements for understanding the user study<br>• No criteria for methodological quality (e.g., sample size, operationalisation, statistical testing) | • Not sufficiently understandable what participants did in the user study or what performance data were analysed |
| Publication types | • Journal articles, conference presentations, book chapters, preprints | • Student theses, dissertations<br>• Non-scientific publications (e.g., blog posts) |
| Language | • English | • All other languages |

A first set of criteria concerns technical factors. The search was restricted to real XAI methods explaining the features considered by real DNN in image classification and related tasks. Studies were excluded if they relied on simulated (X)AI or if they only used types of XAI that are not attribution methods. Attribution methods were included that explain feature relevance using different saliency map visualisations (e.g., heatmaps, image segments, outlines). If extensions of saliency maps were used or other types of XAI were combined with saliency maps, studies were only included if a direct comparison between the extended or combined saliency maps and pure, traditional saliency maps was reported. Moreover, the review focused on image classification and related tasks. Thus, studies were excluded if they used other data than images (e.g., text, videos). The definition of images was restricted to physical objects (e.g., photographs, radiological images), while data visualisations were excluded (e.g., graphs, time series). Image-dependent tasks other than classification were included if they required a selection of relevant image areas (e.g., object detection, visual question answering).

A second set of criteria concerns human factors. Studies were included if they assessed the effects of saliency maps on human performance in two types of tasks: tasks that focus participants' attention on decisions about the image (e.g., classifying bird species, indicating whether a tumour is present) and tasks that focus participants' attention on what the AI is doing (e.g., predicting which class the AI will select, reporting AI biases). The latter have been termed "proxy tasks" in previous work and it has been cautioned against their application (Buçinca et al., 2020). This is because the evaluation outcomes obtained with such tasks can diverge from those obtained with tasks that focus on the decision per se. However, they can be highly informative, for instance regarding the usefulness of saliency maps for AI developers or people involved in safety evaluations of computer vision systems. Therefore, they were included in the present review but task focus served as a major criterion for systematising the empirical outcomes.

Finally, in line with the aim of capturing the diversity of modulating factors, very liberal criteria were set concerning the quality of user studies. First, there were no constraints regarding indirect quality metrics (e.g., number of citations, reputation of journal or conference). Second, the user study was not required to be the main contribution of the publication but could also serve as an add-on to evaluate a technical innovation. Third, user studies were included even if they had methodological deficits (e.g., small sample size, no statistical analyses) as long as the methods were described in a sufficiently understandable manner and did not make the interpretation of the results impossible.

### 4.2 Search process

The literature search was conducted in February 2024. Initially, all studies were sampled that had already been collected during the author's previous work in the field. Second, references from these publications and from previous systematic reviews were retrieved. After having assembled this initial literature collection, the systematic search started. Four databases were searched sequentially: Google Scholar, Web of Science, the ACM Digital Library, and PSYNDEX. The search string consisted of terms from five concept groups that specified the presence of an empirical study with human participants (e.g., "user study"), the focus on XAI (e.g., "explainable"), the type of XAI (e.g., "saliency maps"), the AI context (e.g., "DNN"), and the task domain (e.g., "image classification"). A full list of terms for each concept is provided in Table 2 and the literal search strings for each database are provided in the Appendix (Table A1).

While searching Google Scholar, the full texts were checked immediately and without an Abstract-based preselection step. For the first 200 results, this was done for each publication (unless the author already knew it was irrelevant). For the next 300 results, only publications were checked that seemed like they might be relevant based on their title and highlighted search terms in the text preview. The strategy of immediately checking the full texts was chosen because it turned out to be much more efficient than checking the Abstract. This is because Abstracts often did not indicate that a user study was conducted and almost never indicated whether objective performance was assessed. When searching the other three databases, a more standard Abstract-based approach was chosen because these databases provided much fewer results (given that they do not search the full text), do not show excerpts with highlighted search terms and two of them do not provide access to the full texts.

Following the systematic search, additional publications were retrieved via the snowball principle (i.e., searching neighbouring nodes in the citation graph). For all publications, cited articles were checked if the text suggested they might be relevant. For publications in which the user study was the main contribution, three sources were checked: other publications citing the publication, related publications and other publications of the main authors.

Finally, a non-systematic search was conducted based on different short versions of the search string. To this end, subsets of the search terms were used (e.g., XAI, user study, saliency map) or the inclusion of specific task domains was forced by putting them in inverted commas (e.g., scene classification, scene recognition, object recognition, object detection, fine-grained). This non-systematic search was only performed with Google Scholar because this database does not correctly process some operators of search strings (i.e., ignoring parentheses). It still is a highly valuable source because it relies on a full text search and thus can retrieve relevant publications that the other databases will miss (e.g., if the Abstract does not mention the user study). However, the signal-to-noise ratio is not favourable when considering the full selection of several thousands of publications. Thus, more targeted search strings have the potential to return publications that had remained undetected so far. Table A2 in the Appendix provides an overview of the sources of the final 52 publications.

**Table 2.** Ingredients of the search string, including the relevant concept and terms used to specify it.

| Concepts | Specific terms |
|---|---|
| Empirical study | User study, participants, human subjects, user evaluation, human evaluation, human experiment, empirical study |
| Focus on XAI | XAI, explainable, explainability, interpretability, understandability |
| Type of XAI | Saliency map, heatmap, attention map, attribution map |
| AI context | Neural network, machine learning, DNN, CNN, ML, AI, classifier, black box |
| Task domain | Image, image classification, scene classification, scene recognition, object recognition, object detection, fine-grained |

*Note*. Depending on the requirements of the particular database, variations of these terms were used (e.g., “explainab*” instead of “explainable OR explainability”).

### 4.3 Documentation and coding

User studies that met the inclusion criteria were documented in a table that characterised each study on 70 variables of nine information types: basic description, technological factors, participants and groups, task and design, control conditions, data analysis, detailed results, categorisation of results and meta information. The specific variables are listed in Table 3 and the full documentation table is available at the Open Science Framework (https://osf.io/ax3yd/). If a publication reported more than one relevant user study, each of them was coded separately. Moreover, only information relevant to the present review was extracted. For instance, if participants had to provide additional subjective ratings of the XAI after performing their task, this information was omitted.

**Table 3.** Variables used to document and code the empirical user studies.

| Infomation type | Specific variables |
|---|---|
| Basic description | ID, year, full reference, number of study, domain (images), key findings, unpacked findings |
| Technological factors | DNN, DNN performance, XAI attribution method, XAI attribution method type, extension of XAI attribution method, extension type, XAI appearance, dataset, manipulation to generate biased XAI, variation in images (systematic), essence of image variation, number of images (total), number of relevant trials (total), incorrect trials (percentage) |
| Participants and groups | Type of participants, number of participants (total), number of participants (per group), number of groups |
| Task and design | Type of study, focus of task (AI or image), task assignment, cognitive requirement, procedure, response options, design factors, type of design (within/between/mixed), within factors, within factor levels, between factors, between factor levels, covariates |
| Control conditions | No XAI, versions of same attribution method (qualitatively or quantitatively different), attribution methods, other types of XAI, other saliency maps (non-XAI), no human-AI team |
| Data analysis | Dependent variables, comparison with automated metrics, statistical testing |
| Results (detailed) | XAI effects depend on, XAI benefits, XAI costs, null effects, confounded XAI effects, differences between XAI methods, differences between versions of saliency maps, differences between saliency maps and extensions or combinations, differences between saliency maps and other types of XAI, comparison with automated XAI metrics, other |
| Results (categorically) | XAI benefits, XAI costs, null effects, hypothesised biases, saliency map differences, saliency maps versus extensions combinations or other types of XAI |
| Meta information | Role of user study, publication type, country, source general, source specific, comments on meta information |

## 4.4 Analysis and integration

First, descriptive statistics (i.e., number and percentage of studies) were computed to characterise the included user studies with regard to seven categories: meta information (e.g., year and type of publication, type of user study), technological factors (e.g., DNN model, XAI method), participants (e.g., domain expertise, AI expertise), images (e.g., general datasets like ImageNet, domain-specific images), human tasks (e.g., task focus, task assignment, cognitive requirements), experimental designs (e.g., between or within participants) and data analysis (e.g., dependent variables, reporting of statistical results). Second, an overview of the empirical outcomes was provided. It was analysed how many studies reported benefits, null effects or even costs of saliency maps and what specific biases were observed. The main analysis examined how the effects of saliency maps depended on factors related to the human tasks, AI, XAI, images, human participants and comparisons. These factors were derived from the included studies and their outcomes, instead of being predefined in advance. That is, they do not represent all factors that could potentially play a role, but only those factors about which it is possible to draw conclusions based on the available research. Examining the role of these factors was achieved by first listing each factor that affected the results in at least one study (e.g., AI accuracy). Given this list of factors, the documentation table was searched to identify all studies that reported an influence of the respective factor. Each observed outcome regarding the factor was added to the list as a bullet point (e.g., saliency maps are more useful for incorrect than correct AI) and all studies reporting this outcome were added as references. Afterwards, the collection of outcomes and citations was integrated by clustering similar outcomes, explicitly differentiating them from opposite outcomes and discussing potential reasons for the divergence.

# 5 Results

## 5.1 Description of publications and studies

### *5.1.1 Meta information*

In total, 68 user studies were extracted that were featured in 52 publications. A full list is provided in the Appendix (Table A3), including each study's key message, assigned task, associated cognitive requirement and outcomes. An overview of the publications' meta information is provided in Figure 2A-C. The publications appeared between 2016 and 2024, with most being published after 2020. They consisted of mostly conference papers (57.7%), several journal articles (28.8%) and a few preprints (13.5%). In about half of them, the user study was the main contribution versus an add-on to evaluate a technical innovation (48.1% vs. 51.9%, respectively). Almost all user studies were experiments (95.6%), two were interviews (2.9%) and one was a work sample (1.5%).

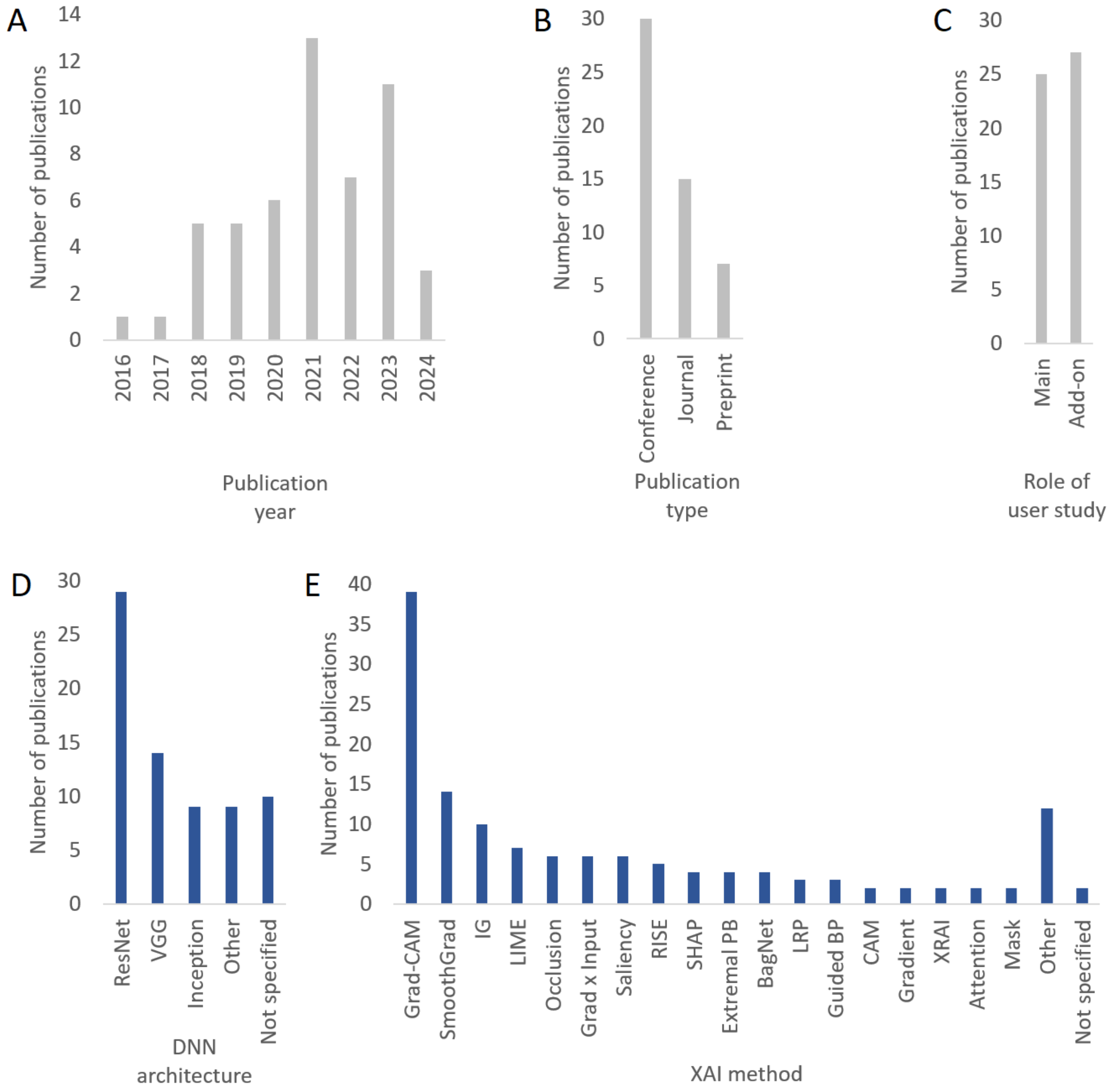

**Figure 2.** Meta information about the publications and technologies. (A) Year of publication, (B) type of publication, (C) role of user study, (D) DNN architecture and (E) XAI method.

#### 5.1.2 Technological factors

Most DNN models were based on some version of the ResNet architecture (42.6%), followed by VGG (20.6%) and Inception (13.2%). Furthermore, 13.2% of the studies relied on other models and 14.7% did not specify the model they used (see Figure 2D). In 23.5% of the studies, the model or dataset was intentionally manipulated to induce biases in the AI predictions or saliency maps.

The application frequencies of XAI methods used to generate saliency maps are presented in Figure 2E. The most frequent XAI method by far was Grad-CAM (57.4%), followed by SmoothGrad (20.6%), Integrated Gradients (14.7%) and the model-agnostic LIME (10.3%). Several other XAI methods were used in fewer than seven studies (< 10%). Exactly half of the studies (50.0%) used more than one XAI method to generate saliency maps. Moreover, about one fourth (29.4%) proposed extensions of standard saliency maps (e.g., showing sets of hierarchically organised saliency maps or combining them with concept-based explanations).

The visual appearance of saliency maps was quite diverse. Although feature relevance was visualised as traditional heatmaps in most of the studies, other visualisations were used as well such as luminance maps, blurring, colour maps, pixel clouds, image segments, outlines and bounding boxes. These saliency maps were either presented next to the fully visible image or directly overlaid on the image. In the latter case, some studies only kept the most relevant areas visible (or invisible when assessing to what degree such masking hampers image recognition).

#### 5.1.3 Participants

At least 61.8% of the studies were conducted with laypeople (mostly recruited via Amazon Mechanical Turk). However, the actual reliance on laypeople presumably was higher. This is because 16.2% of the studies provided no information about their participants but presumably worked with laypeople as well. In 7.4% of the studies, participants had a technical background (e.g., students and researchers in computer science). Moreover, about 15% of the studies were conducted with either of two expert groups: domain experts (8.8%) or AI experts (5.9%).

#### 5.1.4 Images

The majority of studies used images from general datasets like ImageNet that are easy to classify for laypeople (69.1%). These images typically showed a variety of objects but some studies also focused on specific objects categories (e.g., faces, food items, handwritten digits) or presented complex scenes (e.g., landscapes, crowds). In 17.6% of the studies, the images required a fine-grained classification of easily confusable categories (e.g., similar bird species). Other studies used domain-specific images that are typically classified by experts (11.8%). Six used medical images, one used sonar images and one used chess puzzles. Finally, one study did not use any images but only conducted an expert interview.

#### 5.1.5 Human tasks

The human task focused on the AI in 54.4% of the studies, on the image in 39.7% and on both in 5.9%. For each general task focus, the concrete task assignments were coded, describing the actions participants were asked to perform. This coding revealed that AI-focused tasks included predicting AI predictions, predicting AI accuracy or predicting which areas are relevant for the AI, stating one's agreement with the AI prediction, indicating how strongly one would recommend the AI, reporting whether the AI relied on a data artefact and performing different work-related tasks (e.g., reporting of work strategies or challenges encountered when working with saliency maps). Image-focused tasks included simple image classification of everyday objects, difficult fine-grained image classification, image clustering and the selection of chess moves. Table 4 lists all studies for each task focus and task assignment.

**Table 4.** Two task foci with their associated task assignments as well as the numbers and references of the respective studies. N = number of studies.

| Focus | Task assignment | N | Studies |
|---|---|---|---|
| AI | Predicting AI prediction | 21 | (Chandrasekaran et al., 2018; Colin et al., 2022; Fel et al., 2023; Folke et al., 2021; Giulivi et al., 2021; Khadivpour et al., 2022; Kim et al., 2022; Lerman et al., 2021; Ray et al., 2019; Shen & Huang, 2020; Yang et al., 2022; Yang et al., 2021; Zhao & Chan, 2023) |
| | Predicting relevant areas | 3 | (Alqaraawi et al., 2020; Shitole et al., 2021; Sokol & Flach, 2020) |
| | Predicting AI accuracy | 5 | (Alipour et al., 2020; Alqaraawi et al., 2020; Chandrasekaran et al., 2018; Park et al., 2018; Ray et al., 2021) |
| | Agreement with AI prediction | 6 | (Kim et al., 2022; Leemann et al., 2023; Li et al., 2021; Maehigashi et al., 2023a; Stock & Cisse, 2018) |
| | Recommendation of AI | 2 | (Adebayo et al., 2022; Adebayo et al., 2020) |
| | Reporting reliance on artefact | 3 | (Achtibat et al., 2023; Kim et al., 2018; Ribeiro et al., 2016) |
| | Work-related | 3 | (Balayn et al., 2022; Sun et al., 2023) |
| Image | Classification (simple) | 17 | (Biessmann & Refiano, 2019; Cau et al., 2023; John et al., 2021; Kim et al., 2022; Knapič et al., 2021; Leemann et al., 2023; Li et al., 2021; Lu et al., 2021; Maehigashi et al., 2023b; Müller, Thoß, et al., 2024; Nguyen et al., 2021; Schuessler & Weiß, 2019; Selvaraju et al., 2017; Slack et al., 2021; Zhang et al., 2022) |
| | Classification (difficult) | 11 | (Chu et al., 2020; Famiglini et al., 2024; Jin et al., 2024; Jungmann et al., 2023; Kim et al., 2022; Mac Aodha et al., 2018; Nguyen et al., 2021; Richard et al., 2023; Sayres et al., 2019; Wang & Vasconcelos, 2023) |
| | Image clustering | 2 | (Qi et al., 2021) |
| | Chess | 1 | (Puri et al., 2019) |

As it will turn out in Section 5.3.1, it is impossible to understand the task-dependence of XAI effects when only considering the concrete task assignment. This is because one and the same task assignment can yield different outcomes depending on the cognitive requirements associated with it in a given study. Therefore, the main cognitive requirement was coded for each study (e.g., understanding the causes of misclassifications, distinguishing targets from distractors). Figure 3 represents the mapping between task assignments (left-hand side) and cognitive requirements (right-hand side). The figure reveals that a particular task assignment can impose different cognitive requirements and vice versa. For instance, predicting AI predictions (a task assignment) can mean that participants need to detect biases, understand the AI's classification strategies or understand causes of misclassifications, to name but a few. Conversely, bias detection (a cognitive requirement) can be examined by asking participants to predict AI predictions, report whether the AI has relied on a data artefact, state how strongly they would recommend the AI or perform various work-related activities.

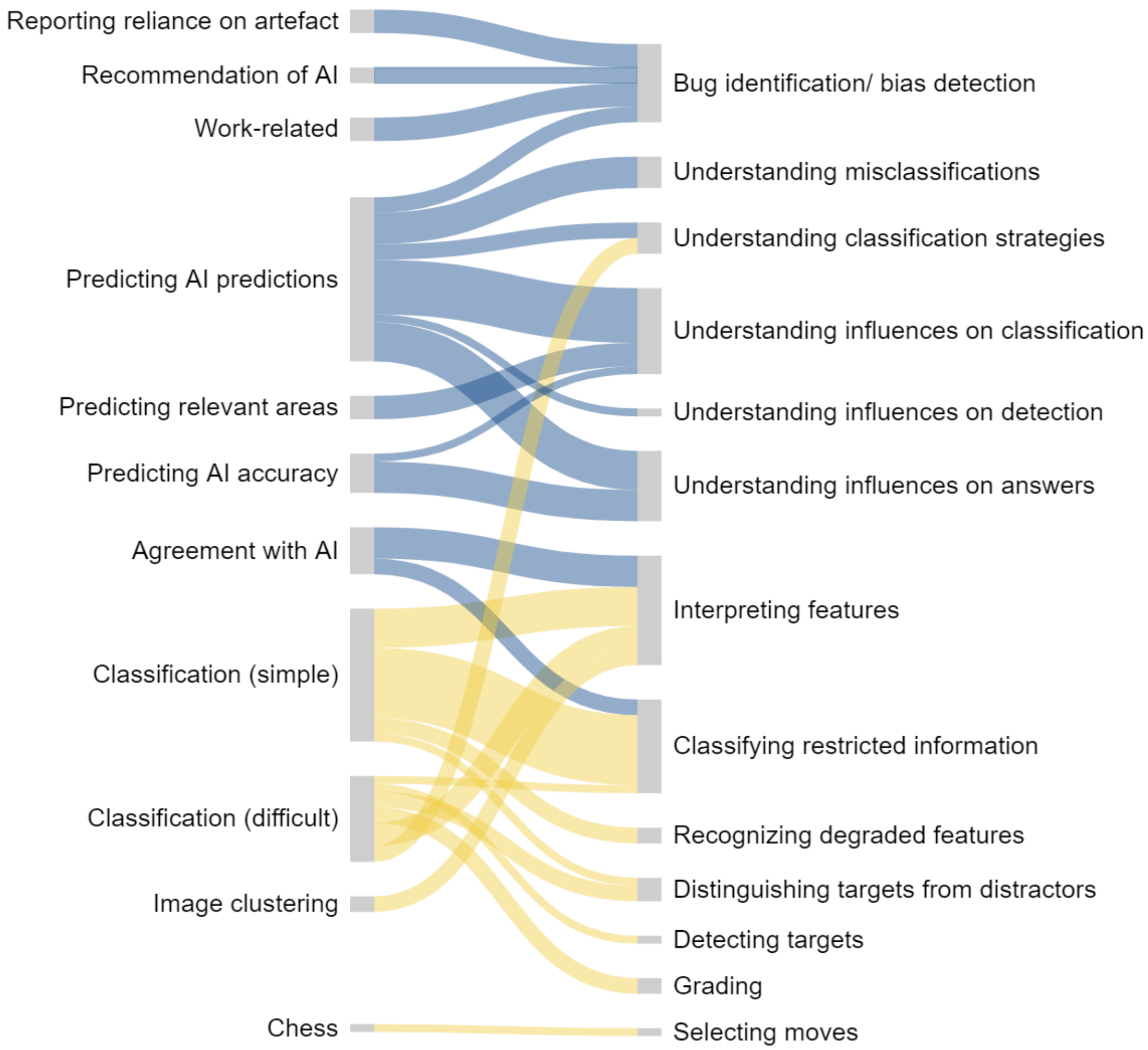

**Figure 3.** Mapping of concrete task assignments (left) to more abstract cognitive requirements (right). AI-focused tasks are presented in blue and image-focused tasks are presented in yellow. The height of the grey bars represents the number of studies. Figure created with SankeyMATIC.com.

*5.1.6 Experimental designs*

Most user studies used mixed designs (47.6%), thus combining factors that were varied within and between participants. Pure between- and within-participants designs were used equally often (22.1%). In 5.9% of the studies, the design was not specified (and could not be inferred from the text) and in 2.9% of the studies, no experimental variation took place. An overview of the factors varied in the studies is provided in Figure 4A. The factor most frequently varied by far was the XAI method (79.4%), followed by AI accuracy (13.2%). Variations of XAI availability are severely underestimated with 11.8%. This is because although 55.9% of the studies used a baseline condition without XAI, most of them directly included it in their XAI method factor when describing their design and analysing their data (e.g., conceptualising the XAI method to have three levels: Grad-CAM, Integrated Gradients, no XAI), instead of coding XAI availability as a separate factor. Other popular factors were image type (14.7%) and the quality of saliency maps (10.3%). Moreover, the temporal factor of training session was varied in studies that investigated whether saliency maps facilitate the learning of fine-grained classification (10.3%). All other factors were varied in less than five studies (< 6%), including the type of AI bug, the quality of saliency maps or the provision of additional information about the AI outcome.

*5.1.7 Data analysis*

Three types of empirical outcomes were reported in the studies (see Figure 4B). Most studies analysed the effects of saliency maps (i.e., benefits, costs, null effects) compared to a baseline without saliency maps (69.1%). Some studies analysed whether different saliency maps led to different outcomes (55.9%) and some analysed whether saliency maps were superior or inferior to extensions and combinations with other types of XAI (36.8%). The dependent variables used to analyse human task performance are summarised in Figure 4C. The majority of studies (77.9%) assessed participants' accuracy in solving the respective task. Fewer studies assessed solution times (14.7%) or the rate of agreement with the AI predictions (13.2%). Moreover, 23.5% of the studies used alternative measures to evaluate human performance. Some of those can be interpreted as proxies for accuracy (e.g., win rate, number of identified bugs), some reflect performance-based confidence measures, behavioural changes or consequences of behaviour. Others were derived from qualitative content analyses of participants' verbal reports. Only 52.9% of the studies reported statistical analyses of their human performance data, including studies that did not provide statistical values but verbally indicated whether comparisons were statistically significant. Moreover, 14 studies (20.6%) compared their human performance data to automated XAI evaluation metrics.

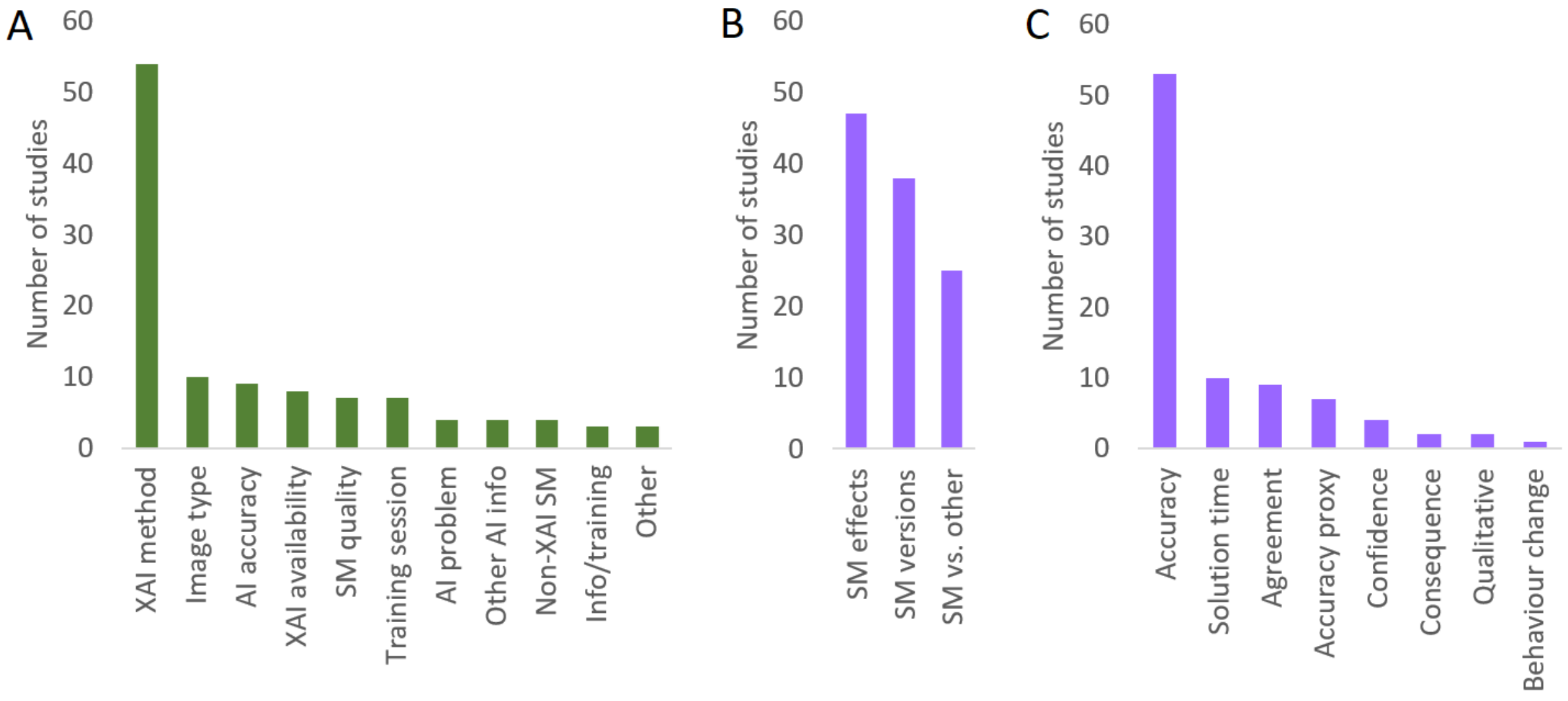

**Figure 4.** Experimental conditions and data analysis. (A) Factors varied in individual studies, (B) type of empirical outcome (with "SM versions" being an abbreviation for "differences between SM versions" and "SM vs. other" being an abbreviation for "differences between SM and other types of XAI") and (C) dependent variables. SM = saliency maps.

## 5.2 Overview of human performance outcomes

*5.2.1 Distribution of effects*

Forty-seven of the 68 studies (69.1%) compared the effects of saliency maps to baselines without XAI. Benefits of saliency maps were found in 55.3% of these studies, costs in 29.8% and null effects in 70.2% (see Figure 5A, top). Note that the values do not add up to 100% as one and the same study may report several outcomes (e.g., for different experimental conditions or dependent variables). Thirty-eight of the 68 studies (55.9%) used more than one version of saliency maps. Differences between these versions were present in 47.4% and absent in 23.7% of the studies (see Figure 5A, middle) and 29.9% did not compare the outcomes between the saliency maps they used. Finally, 25 studies (36.8%) compared saliency maps to extensions or other types of XAI. Saliency maps turned out to be superior in 16.6% of these studies, inferior in 76.0% and conflicting outcomes or null effects were observed in 8.0% (see Figure 5A, bottom). These percentages should not be overinterpreted as they only indicate that an outcome was present in a study, but not how strong or consistent it was. For instance, if a study reported overall benefits of saliency maps but costs in a single condition, it would still contribute to the category of benefits and costs equally. Therefore, a quantitative integration of the results is not particularly informative and the effects will be investigated more thoroughly in the following sections.

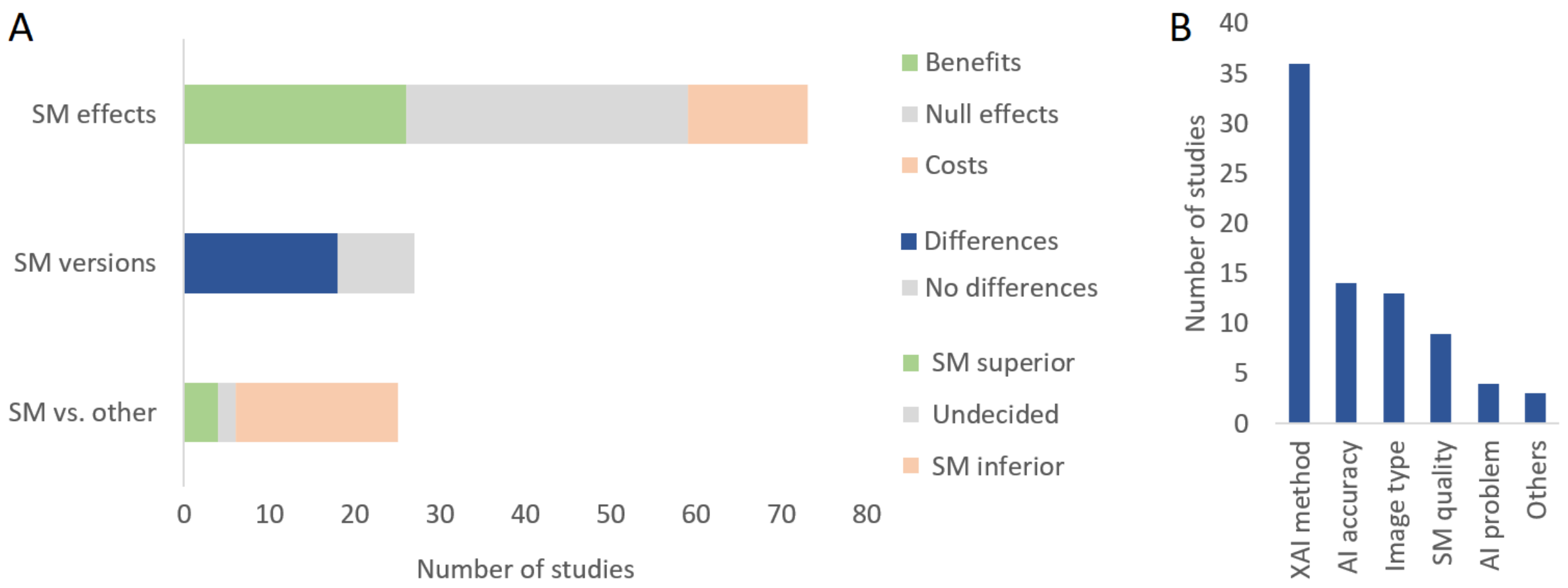

**Figure 5.** Summary of results in terms of the number of studies reporting particular effects on human performance. (A) Effects of saliency maps compared to a baseline without XAI (top), differences between versions of saliency maps (middle) as well as differences between saliency maps and extensions or other types of XAI (bottom). (B) Factors found to modulate the effects of saliency maps within individual studies. SM = saliency maps.

*5.2.2 Biases observed when using saliency maps*

Before delving into the factors that modulated the effects of saliency maps, it may be informative to gain an overview of the reported biases. First, when saliency maps highlight informative areas, people tend to *infer that the AI is correct* (Ray et al., 2021). Consequently, saliency maps can increase human *agreement with AI predictions* without concurrently increasing the accuracy of human performance (Cau et al., 2023; Kim et al., 2022; Maehigashi et al., 2023a; Nguyen et al., 2021; Stock & Cisse, 2018). What makes higher agreement problematic is that saliency maps can lead people to agree with the AI in the wrong situations, making them more likely to agree with unconfident AI predictions and less likely to agree with confident ones (Cau et al., 2023). Moreover, saliency maps can lead people to agree with incorrect AI predictions, making them replicate the AI's errors, particularly when the saliency maps are easy to use (Biessmann & Refiano, 2019; Nguyen et al., 2021; Stock & Cisse, 2018). In this way, even medical experts can be misled into overdiagnosing normal images, hallucinating problems where in fact there are none (Sayres et al., 2019). Conversely, when saliency maps fail to highlight important features, people may miss them as well (Puri et al., 2019).

Another bias is that saliency maps can foster *misinterpretations of feature relevance*. People tend to wrongly assume that the features used in images with correct predictions are relevant, while the features in images with incorrect predictions are irrelevant (Balayn et al., 2022). Moreover, saliency maps do not always enable people to correctly recognise whether the AI has relied on the intended, class-defining image features or on forbidden features like image captions (Kim et al., 2018).

Finally, working with saliency maps can be *effortful*. Human performance may be slowed down compared to conditions without XAI and cognitive overload may occur (Cau et al., 2023; Puri et al., 2019; Richard et al., 2023). These costs are not restricted to laypeople but are also found when saliency maps are used by experts during AI development (Sun et al., 2023). Taken together, ample evidence suggests that the interaction with saliency maps is prone to biases. The following sections will outline how these effects are modulated by contextual factors.

### 5.3 Modulators of the effects of saliency maps on human performance

One way of investigating performance effects is to analyse which factors modulated them in individual studies. The frequencies of these factors are depicted in Figure 5B. The most frequently reported influence was the XAI method. In 52.9% of the studies, different XAI methods (including attribution methods and other types of XAI) induced different human performance outcomes. Other influential factors were AI accuracy (20.6%), image type (19.1%), the quality and visualisation of saliency maps (13.2%) and the specific AI problem (5.9%). All other factors modulated the effects in no more than a single study.

However, only looking at the factors that were influential within studies can cause an overestimation of some influences and an underestimation of others. For instance, the XAI method was varied in the majority of studies, with some including more than two attribution methods and some including other types of XAI. This makes it highly likely to find at least some influence of this factor and bears the risk of substantially overestimating its importance. Conversely, it will turn out in the next section that some of the most profound influences were related to the human tasks. This would remain completely undetected if only searching within studies, because individual studies rarely used more than one task. A systematic review has the potential to mitigate this problem by integrating the empirical outcomes across studies. To this end, the following sections will investigate how human performance was modulated by factors related to the task, AI, XAI, images, humans and comparisons. An overview of these factors is provided in Figure 6.

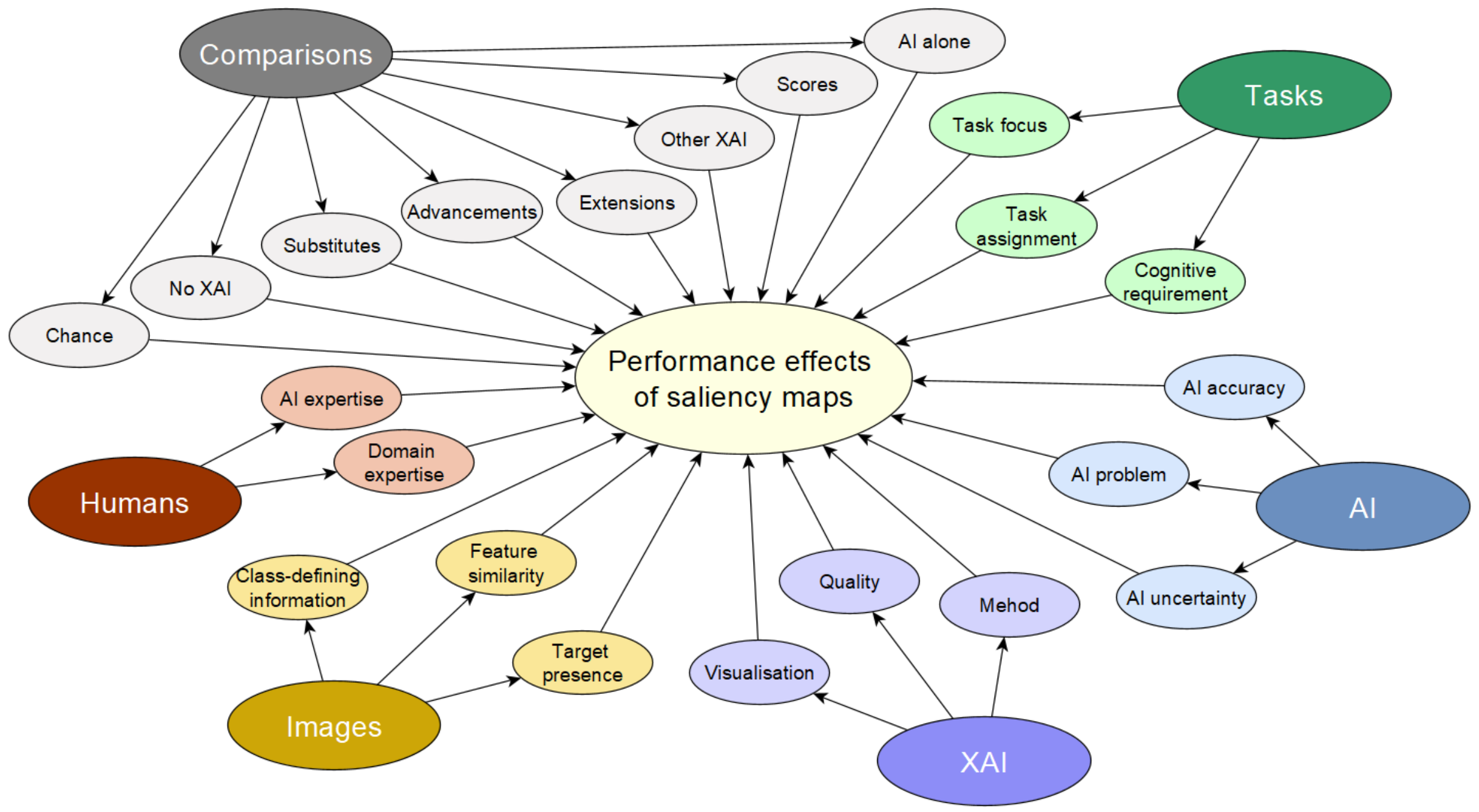

**Figure 6.** Factors that modulate the effects of saliency maps on human performance.

#### *5.3.1 Task-related factors*

A comparison of saliency maps to baselines without XAI can yield benefits, null effects or costs. Figure 7 maps these three types of outcomes to the human tasks employed in the studies. Specifically, it creates a mapping between task foci, task assignments, cognitive requirements and empirical outcomes. At the intersection between cognitive requirements and outcomes, some nodes have more outgoing than ingoing connections. This is because a single study can report different outcomes, either for different experimental conditions or for different dependent variables. The following sections will disentangle this complex figure by discussing the detailed effects of tasks on the effects of saliency maps.

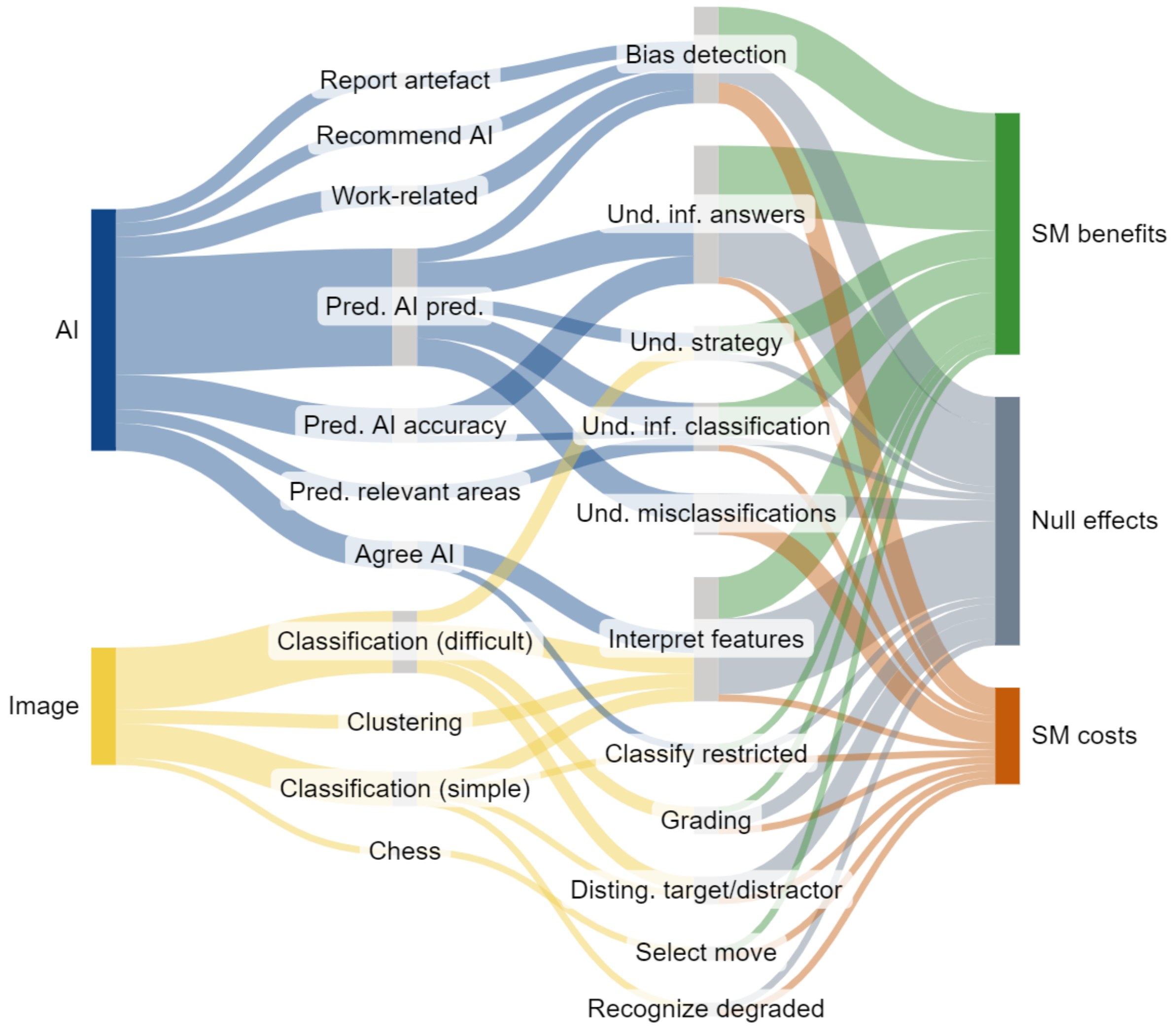

**Figure 7.** Mapping between task foci (left), task assignments (mid-left), cognitive requirements (mid-right) and outcomes (right) for studies that compared saliency maps to a baseline without XAI. The height of the vertical bars represents the number of studies. Figure created with SankeyMATIC.com.

**AI-focused tasks.** When aiming to understand the influence of tasks, an obvious first step is to look at concrete task assignments. How did the effects depend on the activities participants were asked to perform? It will come as no surprise that saliency maps were helpful when participants had to explicitly *identify which image areas affected the AI prediction* (Alqaraawi et al., 2020; Balayn et al., 2022; Lerman et al., 2021; Sokol & Flach, 2020). In fact, this is exactly what saliency maps are made for. Conversely, for other task assignments the findings were less conclusive. When participants' task was to *predict the AI accuracy*, two studies found saliency maps to be helpful (Alqaraawi et al., 2020; Park et al., 2018), whereas two others did not (Chandrasekaran et al., 2018; Ray et al., 2021). Moreover, one study found saliency maps to be helpful for incorrect but not for correct AI predictions (Alipour et al., 2020). A similarly inconclusive picture emerges when the task was to *predict the AI prediction*. Two studies reported that saliency maps were not helpful (Chandrasekaran et al., 2018; Ray et al., 2019) and one even reported that they were harmful (Shen & Huang, 2020). Not a single study reported consistent benefits but several studies found that the effects depended on other factors such as the cognitive requirements underlying the task, AI accuracy or XAI methods (Colin et al., 2022; Fel et al., 2023; Folke et al., 2021; Kim et al., 2022; Lerman et al., 2021; Yang et al., 2022). The role of these factors will be discussed below. For now, it suffices to say that it seems impossible to make general statements about the influence of task assignments. This conclusion is nicely exemplified in an article that reported three studies (Colin et al., 2022). Superficially, they were identical in terms of their task assignment: participants always had to predict AI predictions. However, the studies posed completely different cognitive requirements: bias detection, understanding which strategies the AI uses to accurately perform fine-grained classification and understanding why the AI frequently misclassifies particular images. The authors reported benefits of saliency maps in the first two studies but not in the third one. Thus, it seems informative to scrutinise the role of specific cognitive requirements.

A first cognitive requirement is *bug identification or bias detection*. In this context, the evidence for benefits of saliency maps is most consistent (Achtibat et al., 2023; Adebayo et al., 2022; Adebayo et al., 2020; Balayn et al., 2022; Colin et al., 2022; Fel et al., 2023; Ribeiro et al., 2016; Sun et al., 2023). On the one hand, the usefulness of saliency maps was confirmed in interviews with AI experts (Balayn et al., 2022; Sun et al., 2023). Professional AI developers explained how they worked with saliency maps and what challenges they encountered. Their reports revealed that saliency maps are highly conducive in retaining model generalisability. They can support AI experts in generating and testing hypotheses about the features used by the AI, spotting unreasonable attention, checking the AI's vulnerabilities and assessing why it lacks robustness. On the other hand, the usefulness of saliency maps for bias detection has been confirmed in experimental studies that analysed specific indicators of task performance. Saliency maps made participants more likely to report that the AI relied on a bias (Achtibat et al., 2023; Ribeiro et al., 2016) and thus more likely to correctly predict the incorrect AI prediction in biased images (Colin et al., 2022; Fel et al., 2023). Moreover, when participants saw saliency maps, they were less likely to recommend a biased AI for sale but more likely to recommend a normally functioning AI (Adebayo et al., 2022; Adebayo et al., 2020).

Despite these benefits of saliency maps in bias detection tasks, there are a number of risks associated with such explanations. First, in the two studies just mentioned (Adebayo et al., 2022; Adebayo et al., 2020), participants did not fully reject the biased AI when working with saliency maps, which led the authors to conclude that they were of limited value. Second, saliency maps can induce misinterpretations of feature relevance (Kim et al., 2018): in some cases, they did not enable participants to correctly recognise whether the AI based its classification on the actual image content or on image captions. Moreover, saliency maps can induce confirmation bias in AI experts' hypothesis testing, keeping them from searching for more bugs after having found an apparent problem based on only a few examples (Balayn et al., 2022). Such cherry-picking may leave it unclear whether a detected problem affects the whole dataset and even if it does, the implications for action are not always obvious (Sun et al., 2023). Despite these risks, it can be concluded that saliency maps are a valuable tool to support bias detection.

For other cognitive requirements, the evidence base is more limited. First, saliency maps were found to be helpful when the task required participants to *understand which strategies are used by the AI* to correctly perform difficult fine-grained classification tasks (Colin et al., 2022; Fel et al., 2023). In both studies, the authors reported that with the help of saliency maps, participants could learn to predict which class the AI will select for two highly similar types of leaves. Conversely, saliency maps were consistently found not to be helpful when the task required participants to *understand why the AI misclassifies images* (Colin et al., 2022; Fel et al., 2023; Shen & Huang, 2020). For instance, when the AI frequently confused two similar fox species, saliency maps did not enable participants to learn predicting which of them the AI will select.

An interesting case is tasks that require participants *to understand which image features affect the AI's performance*. While this cognitive requirement seems closely matched to the capabilities of saliency maps, the results are inconsistent. In the context of image classification, saliency maps were helpful in one study (Alqaraawi et al., 2020) but in another study it depended on AI accuracy whether they were helpful or harmful (Folke et al., 2021). Analogous results were reported in the context of visual question answering. One study found positive effects throughout (Park et al., 2018), one found null effects (Ray et al., 2021) and three others found mixed effects depending on other factors such as AI accuracy or saliency map quality (Alipour et al., 2020; Lerman et al., 2021; Ray et al., 2019). Thus, even when the task requires participants to understand the role of particular image features, this is no guarantee that saliency maps will be useful. Still, overall it appears that in AI-focused tasks, saliency maps have the potential to support human performance.

**Image-focused tasks.** A very different picture emerges for image-focused tasks. First, saliency maps turned out not to be helpful when participants had to perform *simple image classification tasks* (Knapič et al., 2021; Nguyen et al., 2021). For instance, they did not increase participants' ability to decide whether everyday objects from ImageNet matched a given label (Nguyen et al., 2021). This might not be too surprising when considering the simplicity of the task. Still, two studies did report benefits in simple classification tasks (Kim et al., 2022; Lu et al., 2021). In the first study, participants saw an image and four predicted classes with their respective saliency maps (Kim et al., 2022). However, generic labels were used (e.g., class 1, class 2), forcing participants to rely on the saliency maps as they were the only thing that differentiated between the four alternatives. Not surprisingly, performance with saliency maps was better than random guessing. Similarly, a chance baseline was also used in the second study that found benefits of saliency maps in a simple classification task (Lu et al., 2021). Participants were better able to classify image segments when they were selected by an actual XAI method than when they were selected randomly. Arguably, the comparison of saliency maps to chance does not provide a very strong test of their quality. Thus, there is no convincing evidence that saliency maps support simple image classification tasks.

However, they do not seem particularly helpful in *difficult classification tasks*, either. This was reported in three types of studies. The first type required laypeople to classify images that were familiar per se but the task was inherently difficult: estimating the exact age of faces (Chu et al., 2020). Performance did not improve when participants saw saliency maps. Presumably, the difficulty of the task did not result from participants not knowing where to look but from not knowing how to translate the observed features (e.g., wrinkles) into exact numbers.

The second type of study required laypeople to perform fine-grained classification of unfamiliar images such as highly similar bird species or dog breeds (Kim et al., 2022; Nguyen et al., 2021; Qi et al., 2021). In one study, saliency maps had no effects overall, while for the very difficult fine-grained classification of adversarial images they even impaired performance (Nguyen et al., 2021). Presumably, when participants did not know the presented objects, saliency maps nudged them to agree with the AI predictions even when they were incorrect. In another study, participants had to cluster images of highly similar birds (Qi et al., 2021). Standard saliency maps did not enable them to perform the clustering like the AI. Benefits of saliency maps were only found in the study that used generic labels and compared performance to random guessing (Kim et al., 2022). Thus, replicating the results for simple classification tasks, saliency maps only supported difficult classification tasks when a very weak baseline was used.

The findings are no more optimistic in the third type of study that used saliency maps in difficult classification tasks: saliency maps did not tend to support domain experts in classifying images from their work domain. In medical image classification, they usually were not helpful compared to baselines without XAI (Jin et al., 2024; Jungmann et al., 2023; Sayres et al., 2019). Analogous findings were reported in another expert domain: saliency maps did not increase experts' classification accuracy for sonar images of mines (Richard et al., 2023). Taken together, using more difficult tasks does not seem to increase the usefulness of saliency maps.

There is one type of image-focused task in which saliency maps were consistently found to be useful: *category learning*. In these studies, saliency maps served as feedback while teaching fine-grained classification strategies. When used in this way during training, they increased performance accuracy in a subsequent test phase (Mac Aodha et al., 2018; Wang & Vasconcelos, 2023). These findings match the ones observed for AI-focused strategy understanding (Colin et al., 2022; Fel et al., 2023).

Given that saliency maps rarely improved human image classification performance, one might wonder whether they are more helpful in *other tasks*. One study examined their potential to support experienced chess players in selecting their moves (Puri et al., 2019). A particular type of saliency map indeed was helpful, while two others even impaired performance. Still, this partial benefit might be instructive as it reflects an instance where the capabilities of saliency maps matched the specific task requirements. When the key question is *where* something is located, saliency maps can provide exactly the information that is needed. Otherwise, they seem to be of limited use.

#### *5.3.2 AI-related factors*

**AI accuracy**. In AI-focused tasks, saliency maps were often found to be *more helpful for incorrect AI predictions* (Alipour et al., 2020; Balayn et al., 2022; Folke et al., 2021; Yang et al., 2022). Similarly, they were helpful to determine that the AI indeed relied on a data artefact but less helpful to determine that it did not (Achtibat et al., 2023). In all of these studies, saliency maps either did not help or even were harmful when the AI was correct. To account for this rather consistent finding, it was suggested that saliency maps help people distinguish between their own and the AI's ease of classification (Yang et al., 2022; Yang et al., 2021). That is, saliency maps may enable people to understand that the AI can err even if they themselves think an image is easy to classify. In line with this hypothesis, saliency maps were particularly helpful when the AI classified familiar images incorrectly, but not when it classified unfamiliar images correctly.

However, there also are studies in which saliency maps *did not support participants in dealing with misclassifications*, either (Colin et al., 2022; Fel et al., 2023; Kim et al., 2022; Shen & Huang, 2020). For one, this was observed when the underlying cognitive requirement was to understand why the AI tends to confuse two highly similar classes (Colin et al., 2022; Fel et al., 2023). Moreover, saliency maps can make people falsely accept incorrect AI predictions (Kim et al., 2022). In all three studies, saliency maps presumably failed to support dealing with incorrect AI predictions because they looked almost identical for correct and incorrect predictions. Thus, it seems like the findings can be traced back to image-related factors (see Section 5.3.4). Finally, one study even reported a benefit for correct AI predictions: saliency maps can help AI developers understand that bugs may exist even when the AI is classifying the images correctly (Balayn et al., 2022). In spite of these exceptions, the majority of the evidence suggests that in AI-focused tasks, saliency maps are more helpful when the AI is incorrect.

The opposite was found for image-focused tasks, where saliency maps only were helpful when the AI was correct. First, only explanations of correct AI predictions helped medical experts determine whether a disease was present or not (Sayres et al., 2019). Second, saliency maps helped novices learn which features the AI was using to correctly perform difficult fine-grained classification tasks (Colin et al., 2022; Fel et al., 2023; Mac Aodha et al., 2018; Wang & Vasconcelos, 2023). The double dissociation between task focus and AI accuracy seems plausible, given the different functions that saliency maps fulfil in these tasks. In AI-focused tasks, they are supposed to help people notice when the AI does something wrong, whereas in image-focused tasks, they are supposed to help people extract the right information to classify the image correctly.

**AI uncertainty**. A factor closely related to AI accuracy is AI uncertainty. Saliency maps can have a paradoxical relation with AI uncertainty, prompting people to contradict the AI in the wrong situations (Cau et al., 2023). That is, saliency maps led people to agree more with unconfident AI predictions but less with confident AI predictions compared to a baseline without XAI.

**Type of AI problem**. Saliency maps are more helpful in dealing with some AI bugs and biases than others. First, their effects depended on the *type of bug* (Adebayo et al., 2022). They induced more realistic evaluations of biased AI models in case of spurious correlations but not in case of labelling errors, re-initialised weights and out-of-distribution errors. In the latter cases, participants tended to ignore the saliency maps and only rely on the labels. Moreover, both the benefits and costs of saliency maps can depend on the *causes of misclassification*. Regarding their benefits, saliency maps facilitated the prediction of incorrect AI predictions more strongly for adversarial attacks than normal misclassifications (Folke et al., 2021). Regarding their costs, saliency maps selectively impaired human performance when misclassifications were due to similar appearances or correlations with background features (Shen & Huang, 2020). Conversely, they produced null effects when misclassifications were due to labelling errors, the AI choosing another object or the AI only attending to parts of the relevant object. Taken together, saliency maps can be helpful when the AI uses image areas other than those intended to define the class.

### *5.3.3 XAI-related factors*

**Attribution method**. Two types of XAI-related influences can be distinguished: those that depend on the attribution method and those that depend on variations in the quality and visualisation of saliency maps while the same method is used. When discussing effects of attribution methods, the present review is not interested in evaluating particular methods (e.g., concluding whether Grad-CAM is superior or inferior to Integrated Gradients). Instead, the aim is to investigate under what conditions different methods are likely to induce differences in human performance. For a detailed inspection of the effects of particular methods, readers are referred to the documentation table provided via the Open Science Framework.

Several studies reported human performance differences depending on attribution methods. These effects were most prominent when participants had to classify image segments that were generated by different attribution methods (Biessmann & Refiano, 2019; John et al., 2021; Lu et al., 2021; Müller, Thoß, et al., 2024; Selvaraju et al., 2017). Obviously, this task depends on the attributions as it requires a classification of highly restricted information. Therefore, it is important to know whether differences between attribution methods were also found in other tasks. In some cases they were. One study required participants to report whether the AI had relied on a data artefact (Achtibat et al., 2023). Despite the high potential of saliency maps to support bias detection in general (see Section 5.3.1), only Grad-CAM was helpful in this task, while performance was at chance with Integrated Gradients. Similar results were found when the task was to recommend the AI for sale (Adebayo et al., 2020). Another example is the selection of chess moves (Puri et al., 2019). Participants strongly benefitted from one method of generating saliency maps, whereas two other methods even impaired performance as they either highlighted too few or too many positions. Moreover, the superiority of attribution methods can depend on other factors such as image type (Müller, Thoß, et al., 2024). When the class was defined by singular objects, participants could more easily classify segments generated by XRAI than Grad-CAM, while the reverse was true for complex scenes.

Next to these observations, there are numerous studies in which human performance did not depend on the attribution method (Chandrasekaran et al., 2018; Khadivpour et al., 2022; Knapič et al., 2021; Nguyen et al., 2021). Moreover, trade-offs may occur when attribution methods are beneficial in one way but problematic in another. For instance, it was suggested that methods which generate more easily usable explanations can be more seductive, making participants replicate the AI's errors (Biessmann & Refiano, 2019).

**Quality of saliency maps**. Several studies manipulated the quality of saliency maps generated by one and the same attribution method. One study assessed this quality based on participants' subjective ratings (Ray et al., 2019). Benefits of saliency maps were only found for saliency maps with high ratings, while those with low ratings even decreased performance. Other studies conceptualised low-quality saliency maps as those that highlight areas considered irrelevant by humans. These saliency maps were often generated by inducing spurious correlations, leading the AI to focus on background features. One might expect that such manipulations strongly modulate the effects of saliency maps. Surprisingly, this is not supported by the available evidence. In fact, the quality of saliency maps had little impact on performance (Chu et al., 2020; Leemann et al., 2023; Maehigashi et al., 2023a). For instance, participants' accuracy in classifying the age of faces did not differ between normal and distorted saliency maps (Chu et al., 2020). Presumably, the impacts of saliency map quality are weak and thus drown in the high variance caused by other factors such as the ambiguity of images (Leemann et al., 2023) or the type of dataset (Maehigashi et al., 2023b). However, not only were distorted saliency maps less detrimental than expected. Well-positioned, interpretable saliency maps can even be disadvantageous, leading participants to agree with AI predictions regardless of whether they are correct or not (Maehigashi et al., 2023a).

**Visualisation of saliency maps**. Instead of varying the positioning of saliency maps, two studies varied their visualisation. In one study, radiologists' accuracy in diagnosing bone fractures depended on two visual features (Famiglini et al., 2024). First, accuracy was higher with fine-grained saliency maps that showed detailed features than with coarse-grained ones that showed broad areas. Second, accuracy was higher when saliency maps were visualised as traditional heatmaps than when they had a semantic red/blue colour-coding depending on whether a fracture was present or absent. However, the impact of visualisation seems small. In the study by Famiglini et al. (2024), each visualisation feature only had effects in a particular participant group. In another study, performance did not depend on whether a heatmap or blur visualisation was used (Yang et al., 2021).

### *5.3.4 Image-related factors*

**Target presence**. Some classification tasks require participants to decide whether a feature of interest is present or not. In a medical context, saliency maps only increased diagnostic accuracy when the disease was actually present (Sayres et al., 2019). In contrast, they were even harmful when the disease was absent, leading physicians to falsely assign diagnoses to normal images.

**Class-defining information**. Different types of information contained in an image may determine the class of the image. Not all of them may be amenable to saliency maps. One study found that people could easily use saliency maps to classify image segments when the class was determined by a singular, localised object (Müller, Thoß, et al., 2024). Conversely, saliency maps were less usable when classification relied on object-to-object relations or global scene properties.

**Feature similarity**. The benefits of saliency maps may depend on the perceptual similarity of relevant features (Colin et al., 2022). Across three experiments, the authors showed that saliency maps became less helpful when the features that distinguished between two classes became more similar.

*5.3.5 Human-related factors*

**Domain expertise.** Very few studies addressed human-related factors in general and only one of them investigated the role of domain expertise (Famiglini et al., 2024). For radiologists with higher expertise, granularity mattered: they benefitted more from detailed than coarse-grained saliency maps. For radiologists with lower expertise, colouring mattered: they benefitted more from traditional heatmaps than from saliency maps with semantic colour-coding. Thus, depending on their level of expertise, people may have different information requirements and thus need different visualisations.

**AI expertise.** Only two studies investigated the role of AI expertise. In a realistic bug identification task, AI experts relied on saliency maps more than less experienced participants (Balayn et al., 2022). However, even AI expertise cannot make saliency maps any more useful when the cognitive task requirements do not match their information capabilities. Accordingly, AI experts did not gain any more benefits from saliency maps than laypeople in a simple classification task (Nguyen et al., 2021).

*5.3.6 Comparison-related factors*

Whether saliency maps are useful is highly dependent on the control conditions they were compared to. Most of the previously discussed studies compared saliency maps to a baseline without XAI. The following sections will focus on studies that relied on other comparisons.

**Chance.** Perhaps the weakest baseline to assess the effects of saliency maps is chance. This either meant that real saliency maps were compared to randomly selected image areas or that participants' performance was compared to random guessing. In some cases, performance with saliency maps was higher than with randomly selected image areas (Lu et al., 2021; Wang & Vasconcelos, 2023), whereas in other cases it was not (Chu et al., 2020). Not surprisingly, performance with saliency maps also tended to be higher than random guessing (Achtibat et al., 2023; Kim et al., 2022; Lerman et al., 2021; Ray et al., 2021). What seems more surprising is that in none of these studies, saliency maps managed to exceed the very liberal chance baseline in all examined conditions. Performance was at chance either for some of the investigated objects (Lerman et al., 2021), for correct AI predictions (Ray et al., 2021), for particular attribution methods (Achtibat et al., 2023) or for particular combinations of attribution method and AI accuracy (Kim et al., 2022).

**XAI substitutes**. A somewhat stronger baseline is XAI substitutes – image areas that look like saliency maps but originate from other meaningful sources. To this end, some studies used bottom-up, image-based saliency that is computed irrespective of any DNN classification (Colin et al., 2022; Fel et al., 2023; Nguyen et al., 2021). With these AI-independent saliency maps, performance in predicting the AI was even worse than without XAI and thus real saliency maps were more helpful (Colin et al., 2022). However, even this can only be expected when the task requirements match the information capabilities of saliency maps. Accordingly, saliency maps were no more helpful than bottom-up image saliency when participants' task was to classify images (Nguyen et al., 2021). Another XAI substitute provides a much stronger baseline: human-generated attention maps. They are often regarded as the ground truth for feature relevance. Therefore, it is inferred that saliency maps are useful if they enable performance that is on par with performance relying on human attention maps. Indeed, participants were able to classify image segments generated by attribution methods as accurately as human-generated segments (Schuessler & Weiß, 2019). However, these effects strongly depend on the attribution method and image type (Müller, Thoß, et al., 2024).

**Confidence scores**. Saliency maps were compared to scores that tell participants how confident the AI is in its prediction (Alipour et al., 2020; Alqaraawi et al., 2020; Cau et al., 2023; Chandrasekaran et al., 2018; Nguyen et al., 2021; Sayres et al., 2019). Despite the obvious benefits of such scores, adding saliency maps has sometimes increased performance (Alipour et al., 2020; Alqaraawi et al., 2020;

Sayres et al., 2019). At other times, adding saliency maps was not helpful (Chandrasekaran et al., 2018; Nguyen et al., 2021; Sayres et al., 2019) or even harmful (Nguyen et al., 2021). Moreover, unlike scores, saliency maps did not allow participants to calibrate their agreement with the AI to its confidence (Cau et al., 2023). Finally, presenting saliency maps on their own did not increase diagnostic accuracy, while presenting them together with AI predictions and scores did (Jungmann et al., 2023).

**Other types of XAI**. Relevance-based saliency maps were compared to concept-based explanations that either selected similar image examples from the dataset or presented prototypes (Cau et al., 2023; Fel et al., 2023; Folke et al., 2021; Khadivpour et al., 2022; Kim et al., 2022; Nguyen et al., 2021; Yang et al., 2021). In most of these comparisons, saliency maps were inferior. This was found when participants had to classify images (Nguyen et al., 2021), understand what features influence classification (Khadivpour et al., 2022) or detect biases (Fel et al., 2023). Analogous results were reported for visual question answering, where saliency maps were less helpful than related questions and answers (Ray et al., 2019). In the few cases where saliency maps were more helpful, this was highly context-dependent. For instance, they enabled more accurate performance than examples when the AI was incorrect but certain (Cau et al., 2023) or incorrect for familiar images (Yang et al., 2021). Conversely, examples were more helpful for correct AI predictions of unfamiliar images. Overall, saliency maps were usually outperformed by other types of XAI.

**Advanced saliency maps**. The term "advanced saliency maps" is used here to refer to saliency maps designed for a specific application, while still showing only one map per image. These specialisations were developed for particular tasks such as object detection (Zhao & Chan, 2023), for particular image types such as faces (John et al., 2021) or for particular image distortions such as privacy blurring (Zhang et al., 2022). Moreover, saliency maps were adapted to contain information about the uncertainty of feature relevance (Slack et al., 2021). All these advanced methods yielded superior performance compared to standard saliency maps.

**Extensions and combinations**. Several studies compared standard saliency maps to extensions and combinations with other types of XAI. These approaches come in three forms: combining multiple saliency maps, integrating saliency maps into larger explainability frameworks and combining them with concept-based explanations.

First, *multiple saliency maps* per image were presented in three ways: as combinations of saliency maps and error maps that show which areas the AI may process incorrectly (Ray et al., 2021), as collections of feature-specific saliency maps (Qi et al., 2021) and as trees of hierarchically organised saliency maps (Shitole et al., 2021; Sokol & Flach, 2020). All these approaches enhanced performance compared to standard saliency maps. For instance, when trees of saliency maps revealed the effects of different image areas on AI confidence, this enabled participants to more accurately predict which areas affect the AI prediction (Shitole et al., 2021).

Second, saliency maps were presented within larger *explanation frameworks*. For instance, they were combined with global explanations in prototypes that aimed to support the evaluation and adaptation of AI models (Balayn et al., 2022; Sun et al., 2023). The contributions of both explanation types were complementary: AI experts relied on global explanations to assess influences on the correct majority, whereas saliency maps were most useful to understand incorrect AI predictions. Other explanation frameworks combined saliency maps with stimulus selection strategies for category teaching (Mac Aodha et al., 2018; Wang & Vasconcelos, 2023) or used multimodal explanations in visual question answering (Alipour et al., 2020; Lerman et al., 2021; Park et al., 2018). In most comparisons, standard saliency maps alone were less helpful than the integrated approaches.

Third, one of the most interesting extensions might be the *combination of saliency maps with concept-based explanations* (Achtibat et al., 2023; Adebayo et al., 2022; Fel et al., 2023; Giulivi et al., 2021; Kim et al., 2018; Qi et al., 2021). In this way, information about *where* the AI has attended is enriched with information about *what* the AI has perceived in the highlighted areas. The latter was achieved by selecting similar image examples from the dataset (Achtibat et al., 2023; Fel et al., 2023), by generating the concept the AI has seen and merging it into the image (Giulivi et al., 2021) or by presenting activation scores for each concept that is present in the image (Adebayo et al., 2022; Kim et al., 2018). Across all studies, this combination of *where* and *what* was more helpful than standard saliency maps.

**AI alone**. It has been argued that a critical control condition is AI performance without any human involvement as this comparison indicates whether XAI increases human-AI collaboration (Fok & Weld, 2023). Usually, performance with saliency maps was no better than the AI alone (Chu et al., 2020; Jin et al., 2024) or the results were mixed (Nguyen et al., 2021; Sayres et al., 2019). In a few studies, saliency maps yielded better performance than the AI alone. However, these results are hard to interpret as it remains unclear how much of the effect can be attributed to saliency maps. This is because similar improvements were also found for control conditions without saliency maps (Sayres et al., 2019) or for distorted and intentionally non-informative saliency maps (Maehigashi et al., 2023b; Nguyen et al., 2021). It has been suggested that XAI only enhances human-AI-collaboration if it helps verify the AI predictions – and that saliency maps are not suitable in this regard (Fok & Weld, 2023). The results of the present review seem to corroborate this conclusion.

## 6 Discussion

It seems appealing to explain the decisions of black-box image classifiers by attributing them to particular image areas. But is it also useful? To answer this question, a systematic literature review integrated the findings of 68 empirical user studies that assessed how saliency maps affect human performance. While some performance benefits were evident, null effects were even more frequent and performance costs were not uncommon. Moreover, characteristic biases were observed. For instance, saliency maps led people to infer that the AI is correct and thus agree with its predictions even when these predictions were incorrect or uncertain. Saliency maps also invited misinterpretations of feature relevance, while not always enabling people to recognise whether the AI has actually relied on the intended features. Additionally, working with saliency maps can be effortful and time-consuming. However, a large variance of outcomes was observed both within and across studies. To explain these inconsistencies, it was examined how the effects were modulated by contextual factors related to the human task, AI, XAI, images, humans and comparison conditions. The following discussion will integrate these dependencies and relate them to previous research, outline perspectives for future research, make the limitations of the present review transparent and derive conclusions about the role of saliency maps in human-AI interaction.

### 6.1 Human tasks are of major importance

The performance effects of saliency maps were strongly modulated by the tasks in which they were used. Benefits were primarily observed in AI-focused tasks. Saliency maps have the potential to help people predict, evaluate and improve AI models. However, these benefits depend on whether the information capabilities of saliency maps are aligned with the cognitive task requirements – that is, whether tasks require people to understand *where* the AI is attending. This alignment is strong, for instance, when people have to detect the AI's biases or understand its successful classification strategies. Conversely, the alignment is weak when people have to predict the AI's misclassification of images with overlapping features. Accordingly, saliency maps were beneficial in the former cases but not in the latter.

A rather bleak picture emerged for image-focused tasks. When people had to classify images (no matter if simple or difficult), the few successes of saliency maps could fully be ascribed to questionable methodological choices or particular experimental conditions. The lack of performance benefits in image-focused tasks resonates with previous findings reported for other types of XAI. It has been concluded that XAI usually does not improve human decision-making (Fok & Weld, 2023; Kandul et al., 2023) and that the findings obtained in AI-focused tasks do not generalise to tasks that focus human attention on the decision per se (Buçinca et al., 2020). The present review extends these conclusions to image classification. It shows that the general findings still hold even though image classification may systematically differ from other tasks, for instance because it allows for a more unambiguous evaluation of AI accuracy and requires fewer inferences beyond the visible data.

Aside from standard image classification, studies that provided convincing evidence for the benefits of saliency maps had one thing in common: successful task performance depended on knowing *where* relevant areas are and which of these localised features should be attended – particularly when the class-defining features were perceptually distinct but hard for humans to identify by themselves. This was most clearly demonstrated for category learning (Mac Aodha et al., 2018; Wang & Vasconcelos, 2023). These observations resonate with previous demonstrations of the power of visual highlights in training attentional expertise (Emhardt et al., 2023; Roads et al., 2016). However, there is reason to assume that saliency maps are not just a valuable training tool. They might even support immediate performance in standard image classification tasks if used in the appropriate contexts. This assumption rests on an extensive literature that demonstrated the benefits of cueing in visual inspection tasks (e.g., Alberdi et al., 2004; Hättenschwiler et al., 2018; Maltz & Shinar, 2003; Yeh & Wickens, 2001). A correct highlighting of relevant image areas has consistently improved human performance. Thus, why were such benefits so rare in the present review? The reason seems to lie in the matching of cognitive task requirements to the information capabilities of saliency maps. In cueing studies, the key cognitive requirement is target detection – a process that primarily relies on information about where important features are located. Unfortunately, in the present review it was not possible to examine whether target detection was supported by saliency maps. This is because the target detection studies had to be eliminated for confounding the benefits of saliency maps with the benefits of AI per se (Cabitza et al., 2022; Cabitza et al., 2023; Natali et al., 2023).

An implication for future user studies of saliency maps is that they should match the cognitive task requirements to the information capabilities of saliency maps. That is, researchers should critically ask whether the task they are intending to use has a fair chance of benefitting from a mere highlighting of image areas. Moreover, it will be an exciting prospect for future research to systematically investigate the role of this match. This could be accomplished by comparing the effects of saliency maps between superficially similar tasks that differ in their cognitive requirements (Colin et al., 2022).

### 6.2 The influence of AI performance is important but task-dependent

A second key factor to influence the effects of saliency maps was AI performance. The direction of these effects depended on the human task. In AI-focused tasks, saliency maps were more helpful when the AI was incorrect. At first glance, there seems to be a paradox regarding AI accuracy. On the one hand, it was suggested that saliency maps help people understand that the AI may err (Yang et al., 2021). On the other hand, it was observed that they do not help them understand why (Colin et al., 2022; Fel et al., 2023; Shen & Huang, 2020). More specifically, the understanding of failure cases (i.e., frequently confused classes) was not supported. This paradox can be resolved when taking a closer look at the types of misclassifications. When misclassifications resulted from dataset bias and spurious correlations, their understanding was indeed supported by saliency maps. In contrast, when they resulted from a high perceptual similarity between competitor classes, saliency maps did not help. Thus, the effects of saliency maps depended on the specific AI problem.

While saliency maps were most helpful for incorrect AI predictions in AI-focused tasks, they only were helpful for correct AI predictions in image-focused tasks. These findings resonate with previous research outside the field of XAI. For target cueing in visual inspection, highlights are known to be helpful when the technical system is correct but detrimental when it is incorrect (e.g., Alberdi et al., 2004; Maltz & Shinar, 2003; Yeh & Wickens, 2001). In the latter case, overreliance on the system makes people miss targets that are present and falsely report targets that are absent. Overreliance is a major problem in human-machine interaction (Mosier & Manzey, 2019; Parasuraman & Manzey, 2010; Parasuraman & Riley, 1997; Parasuraman & Wickens, 2008) and hopes have been voiced that it might be mitigated by XAI. However, XAI can even aggravate this problem: overreliance is among the main types of biases considered in the XAI literature and has been reported in numerous studies (for an overview see Bertrand et al., 2022). This calls for effective interventions to counteract such biases. Several review articles have collected such countermeasures that are largely based on human-machine interaction research (e.g., Bertrand et al., 2022; de Visser et al., 2020). Accordingly, their focus is quite broad and future research should investigate how they might have to be adapted to the context of image classification with saliency maps.

Another perspective for future research is to develop a more nuanced understanding of the effects of AI accuracy. In line with signal detection theory, one can distinguish between situations in which the AI produces correct detections, correct rejections, misses and false alarms. The latter two problems have not usually been distinguished in user studies. This would be relevant, however, because the visual cueing literature suggests that misses and false alarms can have profoundly different effects on human performance (Maltz & Shinar, 2003). Moreover, future research should follow up on initial attempts to classify the reasons for misclassifications (Shen & Huang, 2020). Given that this is done in situations where saliency maps have a fair chance of being useful, insights into the dependence of their effects on problem type can be highly informative.

### 6.3 XAI-related factors were less influential than expected

Factors related to the XAI itself had surprisingly little impact. The choice of attribution methods was most influential when participants had to classify image segments that only retained the areas most relevant to the AI. However, presenting such highly restricted information only is suitable in lab settings to infer subtle differences in saliency map quality. It bears little resemblance to the actual use of saliency maps in real life. Occasionally, the choice of attribution methods also mattered in other tasks, but less than one might expect. This initially puzzling observation is put into perspective when considering the findings regarding saliency map quality. Usually, even major distortions of saliency maps did not reduce human performance. If anything, a higher quality of saliency maps led to overreliance on the AI (Maehigashi et al., 2023a).

This is not to say that it is unimportant or even undesirable to highlight areas that are actually relevant. Rather, it reiterates the importance of matching the cognitive task requirements of user studies to the information capabilities of saliency maps. If saliency maps are irrelevant to the cognitive processes needed for solving the task, their quality will not matter, either. If this match is given, saliency map quality might be crucial. Suggestive evidence stems from a recent study that investigated how simulated XAI affects visual quality control (Müller, Reindel, et al., 2024). As this task required target detection, information about the location of features was essential. Accordingly, misplaced highlights reduced human performance below a baseline without XAI and were more detrimental than any other type of (X)AI error. Similar performance costs of misplaced highlights were observed in other cueing studies (Goh et al., 2005; Yeh & Wickens, 2001). It will be interesting whether such negative effects of distorted highlights will generalise to the field of XAI if the design of user studies allows them to.

### 6.4 Image characteristics are relevant but the evidence is scarce

Image-related factors have received surprisingly little attention to date. In general, saliency maps might be most helpful when the images are easy for humans to classify but the AI is misclassifying them nonetheless. In this context, saliency maps may mitigate biases of belief projection, helping people not to over- or underestimate the abilities of the AI based on how they themselves experience the task (Yang et al., 2021).

Concerning specific image contents, saliency maps appeared to be most helpful when classification relied on individual, localised objects that are actually present in the image and perceptually distinct from competitor classes. In other words, saliency maps might only help people infer what features are used by the AI if this directly follows from where they are located. This again highlights the importance of considering the information capabilities of saliency maps when designing user studies – in this case by critically assessing which image classes can be distinguished from competitors merely based on the location of features.

### 6.5 Influences of human characteristics are poorly understood

Regarding the role of human characteristics, isolated findings suggest that both domain expertise and AI expertise may shape how people use saliency maps. At least partly, the role of expertise might be grounded in a better ability of experts to translate between information about where and what (cf. Famiglini et al., 2024): using highly location-specific visualisations to draw inferences about problem content. However, the exact role of these human factors and their interaction with image factors will have to be elucidated in future studies. Given that expertise influences whether and when people benefit from visual highlights (Alberdi et al., 2008; Hättenschwiler et al., 2018), similar findings might also be obtained with saliency maps when tasks are suitable for investigating such effects.

### 6.6 Comparison conditions are crucial

Finally, the effects of saliency maps depended on the conditions they were compared to. Performance benefits were more likely when the control conditions were weak, but even this was no guarantee. When the control conditions were stronger, they almost consistently outperformed saliency maps. However, it should be noted that user studies of advanced XAI methods are often designed to bring out the benefits of the new method and showcase its superiority over saliency maps, instead of implementing a neutral comparison. Thus, the observed inferiority of saliency maps does not necessarily mean that they would be inferior in every other comparison. Instead, this is likely to depend on the match between XAI information capabilities and cognitive task requirements.

In this regard, a promising avenue for future research is to better understand the specific costs and benefits of combining saliency maps with other types of XAI. What combinations are most helpful in what types of tasks? For instance, consider the combination with either example-based or counterfactual, generative XAI. Under what conditions would it be better to combine saliency maps with (a) XAI methods that select example images from the dataset that the AI considers to be similar to the highlighted area (Fel et al., 2023) or (b) XAI methods that generate visualisations of the object perceived by the AI in the highlighted area (Giulivi et al., 2021)? If these approaches were compared in closely matched tasks, it would even be possible to infer the relative contribution of location-based information and different types of concept-based information to overall human performance.

Moreover, another problem might be solved in this way. If XAI methods and user studies were carefully designed to match cognitive task requirements and XAI capabilities, XAI might actually enhance human-AI collaboration. That is, performance might rise above the level of the AI alone. Currently, this criterion is rarely met in XAI user studies (Fok & Weld, 2023). A similar issue has been observed in human-machine interaction research (Boskemper et al., 2022; Rieger & Manzey, 2024). While there were some benefits due to the human ability to spot occasional system errors, these benefits were more than compensated by the human tendency to contradict and cancel correct system decisions. This tendency to contradict the AI in the wrong situations can be aggravated by saliency maps (Cau et al., 2023). Future research should investigate whether it can be mitigated by combining saliency maps with concept-based XAI that informs users how the AI is interpreting the highlighted features.

## 6.7 Limitations of the present review

### *6.7.1 Selection of user studies*

Several limitations of the present review are worth considering. One set of limitations concerns the selection of user studies. Obviously, there is no guarantee that all relevant studies were retrieved. This is not overly problematic given the current dynamics of the field. Most publications appeared after 2020 and thus it is without question that the current literature collection will no longer be up-to-date in a few months or years. Thus, an exhaustive collection would be temporary, anyway. The author intends to regularly update the literature collection shared via the Open Science Framework. To this end, readers are cordially invited to share any relevant user studies that they are aware of or that they will publish in the future.

Two complementary limitations concern the breadth of study selection, which could have been narrower in some ways and broader in others. First, the review aimed for a large diversity of included studies to enable a nuanced assessment of modulating factors. In consequence, no exclusion criteria were applied to filter out studies due to methodological quality concerns. Such concerns could be raised for the majority of studies, given that most XAI researchers are not trained in the design of human experiments. In fact, even the simplest quality criterion – the reporting of statistical analyses – would have eliminated almost half of the studies. However, many other criteria are at least as important. They concern issues of sample size, operationalisation and experimental design, to name but a few. If the number of publications will keep growing at its current pace, future reviews may focus on user studies that meet high methodological standards. They might even conduct meta-analyses to quantitatively integrate their results.

In other ways, the selection of studies could have been broader. For instance, it would have been interesting to also examine the performance effects of saliency maps for text and video data or for dynamic, interactive games. However, any increase in thematic breadth needs to be compensated for by a reduction in analytical depth. Given that the field of XAI is teeming with broad reviews while more focused reviews are scarce, the present review prioritised depth. Another way of broadening the focus would have been to include studies that relied on simulated XAI. It was decided to exclude them for the sake of generalisability, given that it is hard for humans to simulate the outputs of deep neural networks. However, including them might have increased the share of user studies with high methodological standards. This is because simulated XAI is often used by researchers who are trained in empirical research methods and cognitive science but are unable to technically implement DNN models and XAI methods.

### *6.7.2 Category system*

Another set of limitations concerns the category system used to code and document the user studies. The operationalisation of modulating factors may be ambiguous, which can be illustrated using the example of task-related factors. The ambiguities already start at the highest level: there is a thin line between tasks classified as image-focused versus AI-focused. For instance, consider the following two scenarios. In the first scenario, participants see a cue word and an image segment. They have to classify the segment by indicating whether it matches the word (Müller, Thoß, et al., 2024). In the second scenario, participants see an AI prediction and an image. They have to indicate whether they accept or reject the AI prediction (Maehigashi et al., 2023a). Despite being structurally identical, the former scenario was classified as image-focused and the latter as AI-focused. The distinction was motivated by whether the human's attention is directed to the decision or the AI (cf. Buçinca et al., 2020). However, the boundaries may be fluid, blurring the impacts of task focus on human performance.

A similar argument can be made regarding cognitive task requirements. They were extracted because focusing only on the concrete task assignments was insufficient for understanding the impact of tasks. However, while the task assignment can easily be retrieved from a publication's methods description, the cognitive requirements can only be inferred indirectly, which is prone to interpretations and errors. Moreover, they vary in their specificity. For instance, while "classification based on restricted information" is quite straightforward and distinctive, "interpreting the relevance of features" is a broad category that was applied whenever image classification tasks could not be assigned to a more specific requirement (e.g., target detection, distinguishing targets from distractors, grading). Such ambiguities and inconsistencies in the category system may have affected the results.

At the same time, there are valid alternatives to the present conceptualisation of task-related factors in general and cognitive task requirements in particular. On the one hand, the cognitive requirements could have been defined more abstractly. For instance, one might distinguish XAI systems that aim to support different levels of situation awareness: perception, comprehension and projection into the future (Sanneman & Shah, 2022). However, such abstract distinctions compromise analytical depth. They may be more suitable when aiming to organise broader research fields (e.g., XAI in general) than when aiming to capture nuances within these fields. On the other hand, the cognitive requirements could have been replaced by a more application-oriented distinction. For instance, Davis et al. (2020) identified five use cases for XAI: (1) model debugging and validation, (2) model selection, (3) mental model and model understanding, (4) human machine teaming and (5) model feedback, challenging and prescription. A disadvantage of these broad use cases is that they do not provide deeper insights into the cognitive mechanisms of human-XAI interaction. An advantage is that the findings are easier to transfer into practical applications.

## 6.8 Conclusions

Do saliency maps support human performance? Not in general. However, they can be useful under certain circumstances. This is the case primarily if tasks focus on the AI and require humans to understand that it may err – but only if the causes are location-specific. In image-focused tasks, saliency maps may be useful to understand which localised, non-obvious but perceptually distinctive features differentiate between classes. Instead, if understanding *what* does not directly follow from knowing *where*, saliency maps alone are of little use. In this case, combinations with concept-based methods seem promising. An exciting prospect for future research is to study the detailed mechanisms of how such combinations affect human performance across various task requirements.

## Acknowledgments

The author wants to thank Steffen Seitz, Carsten Knoll, Julian Ullrich and the entire XRAISE project team for valuable discussions of XAI methods and their evaluation. This technical perspective has been highly valuable in gaining a better understanding of the literature.

## Funding

This work was supported by the German Centre for Rail Traffic Research (DZSF) at the Federal Railway Authority within the project "Explainable AI for Railway Safety Evaluations (XRAISE)" and by the German Research Foundation (DFG) under grant number: PA 1232/15-1.

## Declaration of interest statement

The author reports there are no competing interests to declare.

## Data availability statement

The review materials are made available via the Open Science Framework (https://osf.io/ax3yd/). Particularly, this repository provides a table that characterises each of the 68 included user studies with regard to 70 variables.

## Author biography

Romy Müller is a PostDoc at the Chair of Engineering Psychology and Applied Cognitive Research at TU Dresden. Her research focuses on the psychological mechanisms and domain-specificity of human-machine interaction in complex industrial systems. She is particularly interested in understanding and supporting human performance during fault diagnosis and decision-making.

## References

Achtibat, R., Dreyer, M., Eisenbraun, I., Bosse, S., Wiegand, T., Samek, W., & Lapuschkin, S. (2023). From attribution maps to human-understandable explanations through concept relevance propagation. *Nature Machine Intelligence*, *5*(9), 1006-1019. https://doi.org/10.1038/s42256-023-00711-8

Adebayo, J., Muelly, M., Abelson, H., & Kim, B. (2022). Post hoc explanations may be ineffective for detecting unknown spurious correlation. In *International Conference on Learning Representations* (pp. 1-13).

Adebayo, J., Muelly, M., Liccardi, I., & Kim, B. (2020). Debugging tests for model explanations. In *34th Conference on Neural Information Processing Systems* (pp. 1-13), Vancouver, Canada.

Alberdi, E., Povyakalo, A., Strigini, L., & Ayton, P. (2004). Effects of incorrect computer-aided detection (CAD) output on human decision-making in mammography. *Academic Radiology*, *11*(8), 909-918. https://doi.org/10.1016/j.acra.2004.05.012

Alberdi, E., Povyakalo, A. A., Strigini, L., Ayton, P., & Given-Wilson, R. (2008). CAD in mammography: lesion-level versus case-level analysis of the effects of prompts on human decisions. *International Journal of Computer Assisted Radiology and Surgery*, *3*, 115-122. https://doi.org/10.1007/s11548-008-0213-x

Alipour, K., Schulze, J. P., Yao, Y., Ziskind, A., & Burachas, G. (2020). A study on multimodal and interactive explanations for visual question answering. *arXiv preprint*, *arXiv:2003.00431*, 1-10. https://doi.org/10.48550/arXiv.2003.00431

Alqaraawi, A., Schuessler, M., Weiß, P., Costanza, E., & Berthouze, N. (2020). Evaluating saliency map explanations for convolutional neural networks: a user study. In *25th International Conference on Intelligent User Interfaces* (pp. 275-285).

Balayn, A., Rikalo, N., Lofi, C., Yang, J., & Bozzon, A. (2022). How can explainability methods be used to support bug identification in computer vision models? In *2022 CHI Conference on Human Factors in Computing Systems* (pp. 1-16).

Bertrand, A., Belloum, R., Eagan, J. R., & Maxwell, W. (2022). How cognitive biases affect XAI-assisted decision-making: A systematic review. In *Proceedings of the 2022 AAAI/ACM Conference on AI, Ethics, and Society* (pp. 78-91).

Biessmann, F., & Refiano, D. I. (2019). A psychophysics approach for quantitative comparison of interpretable computer vision models. *arXiv preprint*, *arXiv:1912.05011*. https://doi.org/10.48550/arXiv.1912.05011

Boskemper, M. M., Bartlett, M. L., & McCarley, J. S. (2022). Measuring the efficiency of automation-aided performance in a simulated baggage screening task. *Human Factors*, *64*(6), 945-961. https://doi.org/10.1177/0018720820983632

Buçinca, Z., Lin, P., Gajos, K. Z., & Glassman, E. L. (2020). Proxy tasks and subjective measures can be misleading in evaluating explainable ai systems. In *25th International Conference on Intelligent User Interfaces* (pp. 454-464).

Cabitza, F., Campagner, A., Famiglini, L., Gallazzi, E., & La Maida, G. A. (2022). Color shadows (Part I): Exploratory usability evaluation of activation maps in radiological machine learning. In *International Cross-Domain Conference for Machine Learning and Knowledge Extraction* (pp. 31-50), Cham: Springer International Publishing.

Cabitza, F., Campagner, A., Famiglini, L., Natali, C., Caccavella, V., & Gallazzi, E. (2023). Let me think! Investigating the effect of explanations feeding doubts about the AI advice. In *International Cross-Domain Conference for Machine Learning and Knowledge Extraction* (pp. 155-169).

Cau, F. M., Hauptmann, H., Spano, L. D., & Tintarev, N. (2023). Effects of AI and logic-style explanations on users' decisions under different levels of uncertainty. *ACM Transactions on Interactive Intelligent Systems*, *13*(4), 1-42. https://doi.org/10.1145/3588320

Chandrasekaran, A., Prabhu, V., Yadav, D., Chattopadhyay, P., & Parikh, D. (2018). Do explanations make VQA models more predictable to a human? In *Proceedings of the 2018 Conference on Empirical Methods in Natural Language Processing* (pp. 1036-1042).

Chu, E., Roy, D., & Andreas, J. (2020). Are visual explanations useful? A case study in model-in-the-loop prediction. *arXiv preprint*, *arXiv:2007.12248*, 1-18. https://doi.org/10.48550/arXiv.2007.12248

Colin, J., Fel, T., Cadène, R., & Serre, T. (2022). What I cannot predict, i do not understand: A human-centered evaluation framework for explainability methods. In *36th Conference on Neural Information Processing Systems* (pp. 1-14).

Davis, B., Glenski, M., Sealy, W., & Arendt, D. (2020). Measure utility, gain trust: Practical advice for XAI researchers. In *2020 IEEE Workshop on Trust and Expertise in Visual Analytics* (pp. 1-8), Salt Lake City, UT, USA: IEEE.

de Visser, E. J., Peeters, M. M. M., Jung, M. F., Kohn, S., Shaw, T. H., Pak, R., & Neerincx, M. A. (2020). Towards a theory of longitudinal trust calibration in human–robot teams. *International Journal of Social Robotics*, *12*(2), 459-478. https://doi.org/10.1007/s12369-019-00596-x

Eldrandaly, K. A., Abdel-Basset, M., Ibrahim, M., & Abdel-Aziz, N. M. (2023). Explainable and secure artificial intelligence: Taxonomy, cases of study, learned lessons, challenges and future directions. *Enterprise Information Systems*, *17*(9), 2098537. https://doi.org/10.1080/17517575.2022.2098537

Emhardt, S. N., Kok, E., van Gog, T., Brandt-Gruwel, S., van Marlen, T., & Jarodzka, H. (2023). Visualizing a task performer's gaze to foster observers' performance and learning—a systematic literature review on eye movement modeling examples. *Educational Psychology Review*, *35*(1), 23. https://doi.org/10.1007/s10648-023-09731-7

Famiglini, L., Campagner, A., Barandas, M., La Maida, G. A., Gallazzi, E., & Cabitza, F. (2024). Evidence-based XAI: An empirical approach to design more effective and explainable decision support systems. *Computers in Biology and Medicine*, *170*, 108042. https://doi.org/10.1016/j.compbiomed.2024.108042

Fel, T., Picard, A., Bethune, L., Boissin, T., Vigouroux, D., Colin, J., Cadène, R., & Serre, T. (2023). CRAFT: Concept recursive activation factorization for explainability. In *Proceedings of the IEEE/CVF Conference on Computer Vision and Pattern Recognition* (pp. 2711-2721).

Fok, R., & Weld, D. S. (2023). In search of verifiability: Explanations rarely enable complementary performance in AI-advised decision making. *arXiv preprint*, *arXiv:2305.07722*, 1-12. https://doi.org/10.48550/arXiv.2305.07722

Folke, T., Li, Z., Sojitra, R. B., Yang, S. C. H., & Shafto, P. (2021). Explainable AI for natural adversarial images. *arXiv preprint*, *arXiv:2106.09106*, 1-11. https://doi.org/10.48550/arXiv.2106.09106

Giulivi, L., Carman, M. J., & Boracchi, G. (2021). Perception visualization: Seeing through the eyes of a DNN. *arXiv preprint*, *arXiv:2204.09920*, 1-13. https://doi.org/10.48550/arXiv.2204.09920

Goh, J., Wiegmann, D. A., & Madhavan, P. (2005). Effects of automation failure in a luggage screening task: a comparison between direct and indirect cueing. In *Proceedings of the Human Factors and Ergonomics Society Annual Meeting* (pp. 492-496), Los Angeles, CA: SAGE Publications.

Haque, A. K. M. B., Islam, A. N., & Mikalef, P. (2023). Explainable Artificial Intelligence (XAI) from a user perspective: A synthesis of prior literature and problematizing avenues for future research. *Technological Forecasting and Social Change*, *186*, 122120. https://doi.org/10.1016/j.techfore.2022.122120

Hättenschwiler, N., Sterchi, Y., Mendes, M., & Schwaninger, A. (2018). Automation in airport security X-ray screening of cabin baggage: Examining benefits and possible implementations of automated explosives detection. *Applied Ergonomics*, *72*, 58-68. https://doi.org/10.1016/j.apergo.2018.05.003

Jin, W., Fatehi, M., Guo, R., & Hamarneh, G. (2024). Evaluating the clinical utility of artificial intelligence assistance and its explanation on the glioma grading task. *Artificial Intelligence in Medicine*, *148*, 102751. https://doi.org/10.1016/j.artmed.2023.102751

John, T. A., Balasubramanian, V. N., & Jawahar, C. V. (2021). Canonical saliency maps: Decoding deep face models. *IEEE Transactions on Biometrics, Behavior, and Identity Science*, *3*(4), 561-572. https://doi.org/10.1109/TBIOM.2021.3120758

Jungmann, F., Ziegelmayer, S., Lohoefer, F. K., Metz, S., Müller-Leisse, C., Englmaier, M., Makowski, M. R., Kaissis, G. A., & Braren, R. F. (2023). Algorithmic transparency and interpretability measures improve radiologists' performance in BI-RADS 4 classification. *European Radiology,*, *33*(3), 1844-1851. https://doi.org/10.1007/s00330-022-09165-9

Kandul, S., Micheli, V., Beck, J., Kneer, M., Burri, T., Fleuret, F., & Christen, M. (2023). Explainable AI: A review of the empirical literature. *Social Science Research Network (SSRN)*, *4325219*, 1-20. https://doi.org/10.2139/ssrn.4325219

Kapishnikov, A., Bolukbasi, T., Viégas, F., & Terry, M. (2019). XRAI: Better attributions through regions. In *Proceedings of the IEEE/CVF International Conference on Computer Vision* (pp. 4948-4957).

Khadivpour, F., Banerjee, A., & Guzdial, M. (2022). Responsibility: An example-based explainable AI approach via training process inspection. *arXiv preprint*, *arXiv:2209.03433*, 1-14. https://doi.org/10.48550/arXiv.2209.03433

Kim, B., Wattenberg, M., Gilmer, J., Cai, C., Wexler, J., & Viegas, F. (2018). Interpretability beyond feature attribution: Quantitative testing with concept activation vectors (TCAV). In *International Conference on Machine Learning* (pp. 2668-2677).

Kim, S. S., Meister, N., Ramaswamy, V. V., Fong, R., & Russakovsky, O. (2022). HIVE: Evaluating the human interpretability of visual explanations. In *European Conference on Computer Vision* (pp. 280-298), Cham: Springer Nature Switzerland.

Knapič, S., Malhi, A., Saluja, R., & Främling, K. (2021). Explainable artificial intelligence for human decision support system in the medical domain. *Machine Learning and Knowledge Extraction*, *3*(3), 740-770. https://doi.org/10.3390/make3030037

Laato, S., Tiainen, M., Islam, A. N., & Mäntymäki, M. (2022). How to explain AI systems to end users: A systematic literature review and research agenda. *Internet Research*, *32*(7), 1-31. https://doi.org/10.1108/INTR-08-2021-0600

Leemann, T., Rong, Y., Nguyen, T. T., Kasneci, E., & Kasneci, G. (2023). Caution to the exemplars: On the intriguing effects of example choice on human trust in XAI. In *37th Annual Conference on Neural Information Processing Systems* (pp. 1-12).

Lerman, S., Venuto, C., Kautz, H., & Xu, C. (2021). Explaining local, global, and higher-order interactions in deep learning. In *IEEE/CVF International Conference on Computer Vision* (pp. 1224-1233).

Li, J., Lin, D., Wang, Y., Xu, G., & Ding, C. (2021). Towards a reliable evaluation of local interpretation methods. *Applied Sciences*, *11*(6), 2732. https://doi.org/10.3390/app11062732

Lu, X., Tolmachev, A., Yamamoto, T., Takeuchi, K., Okajima, S., Takebayashi, T., Maruhashi, K., & Kashima, H. (2021). Crowdsourcing evaluation of saliency-based XAI methods. In *Joint European Conference on Machine Learning and Knowledge Discovery in Databases* (pp. 431-446), Bilbao, Spain: Springer International Publishing.

Mac Aodha, O., Su, S., Chen, Y., Perona, P., & Yue, Y. (2018). Teaching categories to human learners with visual explanations. In *IEEE Conference on Computer Vision and Pattern Recognition* (pp. 3820-3828), Salt Lake City, UT: IEEE.

Maehigashi, A., Fukuchi, Y., & Yamada, S. (2023a). Experimental investigation of human acceptance of AI suggestions with heatmap and pointing-based XAI. In *Proceedings of the 11th International Conference on Human-Agent Interaction* (pp. 291-298).

Maehigashi, A., Fukuchi, Y., & Yamada, S. (2023b). Modeling reliance on XAI indicating its purpose and attention. In *Proceedings of the 45th Annual Conference of the Cognitive Science Society* (pp. 1929-1936).

Maltz, M., & Shinar, D. (2003). New alternative methods of analyzing human behavior in cued target acquisition. *Human Factors*, *45*(2), 281-295. https://doi.org/10.1518/hfes.45.2.281.27239

Mohseni, S., Zarei, N., & Ragan, E. D. (2021). A multidisciplinary survey and framework for design and evaluation of explainable AI systems. *ACM Transactions on Interactive Intelligent Systems*, *11*(3-4), 1-45. https://doi.org/10.1145/3387166

Mosier, K. L., & Manzey, D. (2019). Humans and automated decision aids: A match made in heaven? In M. Mouloua & P. A. Hancock (Eds.), *Human Performance in Automated and Autonomous Systems* (pp. 19-42). CRC Press. https://doi.org/10.1201/9780429458330-2

Müller, R., Reindel, D. F., & Stadtfeldt, Y. D. (2024). The benefits and costs of explainable artificial intelligence in visual quality control: Evidence from fault detection performance and eye movements. *Human Factors and Ergonomics in Manufacturing & Service Industries*, 1-21. https://doi.org/10.1002/hfm.21032

Müller, R., Thoß, M., Ullrich, J., Seitz, S., & Knoll, C. (2024). Interpretability is in the eye of the beholder: Human versus artificial classification of image segments generated by humans versus XAI. *International Journal of Human-Computer Interaction*, 1-24. https://doi.org/10.1080/10447318.2024.2323263

Natali, C., Famiglini, L., Campagner, A., La Maida, G. A., Gallazzi, E., & Cabitza, F. (2023). Color Shadows 2: Assessing the impact of XAI on diagnostic decision-making. In *World Conference on Explainable Artificial Intelligence* (pp. 618-629).

Nguyen, G., Kim, D., & Nguyen, A. (2021). The effectiveness of feature attribution methods and its correlation with automatic evaluation scores. In *35th Conference on Neural Information Processing Systems* (pp. 1-15).

Parasuraman, R., & Manzey, D. (2010). Complacency and bias in human use of automation: an attentional integration. *Human Factors*, *52*(3), 381-410. https://doi.org/10.1177/0018720810376055

Parasuraman, R., & Riley, V. (1997). Humans and automation: Use, misuse, disuse, abuse. *Human Factors*, *39*(2), 230-253. https://doi.org/10.1518/001872097778543886
Parasuraman, R., & Wickens, C. D. (2008). Humans: still vital after all these years of automation. *Human Factors*, *50*(3), 511-520. https://doi.org/10.1518/001872008X312198
Park, D. H., Hendricks, L. A., Akata, Z., Rohrbach, A., Schiele, B., Darrell, T., & Rohrbach, M. (2018). Multimodal explanations: Justifying decisions and pointing to the evidence. In *Proceedings of the IEEE Conference on Computer Vision and Pattern Recognition* (pp. 8779-8788).
Puri, N., Verma, S., Gupta, P., Kayastha, D., Deshmukh, S., Krishnamurthy, B., & Singh, S. (2019). Explain your move: Understanding agent actions using specific and relevant feature attribution. In *Eighth International Conference on Learning Representations* (pp. 1-14).
Qi, Z., Khorram, S., & Fuxin, L. (2021). Embedding deep networks into visual explanations. *Artificial Intelligence*, *292*, 103435. https://doi.org/10.1016/j.artint.2020.103435
Ray, A., Cogswell, M., Lin, X., Alipour, K., Divakaran, A., Yao, Y., & Burachas, G. (2021). Generating and evaluating explanations of attended and error-inducing input regions for VQA models. *Applied AI Letters*, *2*(4), e51. https://doi.org/10.1002/ail2.51
Ray, A., Yao, Y., Kumar, R., Divakaran, A., & Burachas, G. (2019). Can you explain that? Lucid explanations help human-AI collaborative image retrieval. In *Proceedings of the AAAI Conference on Human Computation and Crowdsourcing* (pp. 153-161), Stevenson, WA: AAAI.
Ribeiro, M. T., Singh, S., & Guestrin, C. (2016). Why should I trust you?: Explaining the predictions of any classifier. In *Proceedings of the 22nd ACM SIGKDD International Conference on Knowledge Discovery and Data Mining* (pp. 1135-1144), San Francisco, CA: ACM.
Richard, G., Habonneau, J., Gueriot, D., & Le Caillec, J. M. (2023). AI explainibility and acceptance; a case study for underwater mine hunting. *ACM Journal of Data and Information Quality*, 1-21. https://doi.org/10.1145/3635113
Rieger, T., & Manzey, D. (2024). Understanding the impact of time pressure and automation support in a visual search task. *Human Factors*, *66*(3), 770-786. https://doi.org/10.1177/00187208221111236
Roads, B., Mozer, M. C., & Busey, T. A. (2016). Using highlighting to train attentional expertise. *PLoS ONE*, *11*(2), e0149368. https://doi.org/10.1371/journal.pone.0146266
Rong, Y., Leemann, T., Nguyen, T. T., Fiedler, L., Qian, P., Unhelkar, V., Seidel, T., Kasneci, G., & Kasneci, E. (2023). Towards human-centered explainable ai: A survey of user studies for model explanations. *IEEE Transactions on Pattern Analysis and Machine Intelligence*, 1-20. https://doi.org/10.1109/TPAMI.2023.3331846
Sanneman, L., & Shah, J. A. (2022). The situation awareness framework for explainable AI (SAFE-AI) and human factors considerations for XAI systems. *International Journal of Human–Computer Interaction*, *38*(18-20), 1772-1788. https://doi.org/10.1080/10447318.2022.2081282
Sayres, R., Taly, A., Rahimy, E., Blumer, K., Coz, D., Hammel, N., Krause, J., Narayanaswamy, A., Rastegar, Z., Wu, D., Xu, S., Barb, S., Joseph, A., Shumski, M., Smith, J., Sood, A. B., Corrado, G. S., Peng, L., & Webster, D. R. (2019). Using a deep learning algorithm and integrated gradients explanation to assist grading for diabetic retinopathy. *Ophthalmology*, *126*(4), 552-564. https://doi.org/10.1016/j.ophtha.2018.11.016
Scharowski, N., Perrig, S. A., von Felten, N., & Brühlmann, F. (2022). Trust and reliance in XAI--Distinguishing between attitudinal and behavioral measures. *arXiv preprint*, *arXiv:2203.12318*, 1-6. https://doi.org/10.48550/arXiv.2203.12318
Schemmer, M., Hemmer, P., Nitsche, M., Kühl, N., & Vössing, M. (2022). A meta-analysis of the utility of explainable artificial intelligence in human-AI decision-making. In *Proceedings of the 2022 AAAI/ACM Conference on AI, Ethics, and Society* (pp. 617-626).
Schuessler, M., & Weiß, P. (2019). Minimalistic explanations: capturing the essence of decisions. In *2019 CHI Conference on Human Factors in Computing Systems* (pp. 1-6).

Selvaraju, R. R., Cogswell, M., Das, A., Vedantam, R., Parikh, D., & Batra, D. (2017). Grad-CAM: Visual explanations from deep networks via gradient-based localization. In *IEEE International Conference on Computer Vision* (pp. 618-626), Venice, Italy: IEEE.

Shen, H., & Huang, T. H. (2020). How useful are the machine-generated interpretations to general users? A human evaluation on guessing the incorrectly predicted labels. In *AAAI Conference on Human Computation and Crowdsourcing* (pp. 168-172).

Shitole, V., Li, F., Kahng, M., Tadepalli, P., & Fern, A. (2021). One explanation is not enough: Structured attention graphs for image classification. In *35th Conference on Neural Information Processing Systems* (pp. 1-12).

Slack, D., Hilgard, A., Singh, S., & Lakkaraju, H. (2021). Reliable post hoc explanations: Modeling uncertainty in explainability. In *35th Conference on Neural Information Processing Systems* (pp. 9391-9404).

Sokol, K., & Flach, P. (2020). LIMEtree: Interactively customisable explanations based on local surrogate multi-output regression trees. *arXiv preprint*, *arXiv:2005.01427*, 1-46. https://doi.org/10.48550/arXiv.2005.01427

Stock, P., & Cisse, M. (2018). Convnets and imagenet beyond accuracy: Understanding mistakes and uncovering biases. In *Proceedings of the European Conference on Computer Vision* (pp. 498-512).

Sun, T. S., Gao, Y., Khaladkar, S., Liu, S., Zhao, L., Kim, Y. H., & Hong, S. R. (2023). Designing a direct feedback loop between humans and Convolutional Neural Networks through local explanations. In *Proceedings of the ACM on Human-Computer Interaction* (pp. 1-32).

Sundararajan, M., Taly, A., & Yan, Q. (2017). Axiomatic attribution for deep networks. In *Proceedings of the 34th International Conference on Machine Learning* (pp. 3319-3328).

Vilone, G., & Longo, L. (2021). Notions of explainability and evaluation approaches for explainable artificial intelligence. *Information Fusion*, *76*, 89-106. https://doi.org/10.1016/j.inffus.2021.05.009

Wang, P., & Vasconcelos, N. (2023). A generalized explanation framework for visualization of deep learning model predictions. *IEEE Transactions on Pattern Analysis and Machine Intelligence*, *45*(8), 9265-9283. https://doi.org/10.1109/TPAMI.2023.3241106

Yang, S. C.-H., Folke, N. E. T., & Shafto, P. (2022). A psychological theory of explainability. In *International Conference on Machine Learning* (pp. 25007-25021): PMLR.

Yang, S. C.-H., Vong, W. K., Sojitra, R. B., Folke, T., & Shafto, P. (2021). Mitigating belief projection in explainable artificial intelligence via Bayesian teaching. *Scientific Reports*, *11*(1), 9863. https://doi.org/10.1038/s41598-021-89267-4

Yeh, M., & Wickens, C. D. (2001). Display signaling in augmented reality: Effects of cue reliability and image realism on attention allocation and trust calibration. *Human Factors*, *43*(3), 355-365. https://doi.org/10.1518/001872001775898269

Zhang, W., Dimiccoli, M., & Lim, B. Y. (2022). Debiased-CAM to mitigate image perturbations with faithful visual explanations of machine learning. In *Proceedings of the 2022 CHI Conference on Human Factors in Computing Systems* (pp. 1-32).

Zhao, C., & Chan, A. B. (2023). ODAM: Gradient-based instance-specific visual explanations for object detection. In *11th International Conference on Learning Representations* (pp. 1-30), Kigali, Rwanda.

# Appendix

**Table A1.** Search strings for different databases.

| Database | Search string | N |
|---|---|---|
| Google Scholar | ("user study" OR "participants" OR "human subjects" OR "user evaluation" OR "human evaluation" OR "human experiment") AND (XAI OR explainable OR explainability) AND (saliency map OR heatmap OR attention map OR attribution map) AND "image classification" | 3,720 |
| Web of Science | ((((ALL=("user stud*" OR participant* OR "human subject*" OR "user evaluation*" OR "human evaluation*" OR "human experiment*" OR "empirical stud*")) AND ALL=(XAI OR explain* OR explanat* OR interpretab* OR understandab* OR attribution*)) AND ALL=(saliency OR heatmap OR attention OR attribution)) AND ALL=("neural network*" OR "machine learning" OR DNN OR CNN OR ML OR AI OR classifier OR "black box" OR "black-box")) AND ALL=(image) | 106 |
| ACM Digital Library | [[Abstract: "user study"] OR [Abstract: "user studies"] OR [Abstract: participant*] OR [Abstract: "human subjects"] OR [Abstract: "user evaluation"] OR [Abstract: "human evaluation"] OR [Abstract: "human experiment"] OR [Abstract: "empirical study"]] AND [[Abstract: xai] OR [Abstract: explain*] OR [Abstract: explanat*] OR [Abstract: interpretab*] OR [Abstract: understand*] OR [Abstract: attribution]] AND [[Abstract: saliency] OR [Abstract: heatmap] OR [Abstract: attention] OR [Abstract: attribution]] AND [[Abstract: "neural network"] OR [Abstract: "machine learning"] OR [Abstract: dnn] OR [Abstract: cnn] OR [Abstract: ml] OR [Abstract: ai] OR [Abstract: classifier] OR [Abstract: "black box"] OR [Abstract: "black-box"]] AND [Abstract: image] | 32 |
| PSYNDEX | (DB=PSYNDEX) (XAI OR explain* OR explanat* OR interpretab* OR understandab* OR attribution*) AND (saliency OR heatmap OR attention OR attribution) AND ("neural network*" OR "machine learning" OR DNN OR CNN OR ML OR AI OR classifier) | 28 |

*Note*. N = number of retrieved publications.

**Table A2.** Steps in the search process and number of relevant publications identified in each step.

| Source | N |
|---|---|
| Own work | 1 |
| Previous search | 8 |
| Systematic search (Google Scholar) | 12 |
| Systematic search (Web of Science) | 2 |
| Systematic search (ACM Digital Library) | 0 |
| Systematic search (PSYNDEX) | 0 |
| Non-systematic search | 3 |
| Publications cited by a retrieved publication | 16 |
| Publications citing a retrieved publication | 6 |
| Related work of retrieved publication | 3 |
| Search for specific author | 1 |
| Total | 52 |

*Note*. N = number of relevant publications. Only the first source was coded for each publication (e.g., if a publication already was available from a previous search, it was not counted again when it also appeared in the systematic search results).

**Table A3.** Collection of all studies included in the review.

| Publication | Stu-dy | Task focus | Task assignment | Cognitive requirement | Key findings | Influences on SM effects | SM + | SM − | SM 0 | SM V | SM E |
|---|---|---|---|---|---|---|---|---|---|---|---|
| (Achtibat et al., 2023) | | AI | reporting reliance on data artefact | bug identification/ bias detection | SM can be misleading in the discovery of data artefacts, their benefits depend on the AI outcome, the combination of where + what is most helpful | XAI method, AI accuracy | | | | ✓ | − |
| (Adebayo et al., 2022) | | AI | recommendation of AI | bug identification/ bias detection | SM support the detection of spurious correlations, but only under some conditions, concept-based explanations are more helpful | XAI method, type of AI bug | ✓ | X | ✓ | | − |
| (Adebayo et al., 2020) | | AI | recommendation of AI | bug identification/ bias detection | SM improve evaluations of the AI for spurious correlations, but have relatively weak effects | XAI method, type of AI bug | ✓ | X | ✓ | ✓ | |
| (Alipour et al., 2020) | | AI | predicting AI accuracy (VQA) | understanding what features influence answers | SM only support the prediction of AI accuracy when the AI is wrong | AI accuracy | ✓ | X | ✓ | | + |
| (Alqaraawi et al., 2020) | | AI | predicting relevant areas, predicting AI accuracy | understanding what features influence classification | SM support understanding which features the AI uses, but not how this affects AI predictions | | ✓ | X | X | | |
| (Balayn et al., 2022) | | AI | reporting of work strategies | bug identification/ bias detection | SM support the understanding of AI problems, particularly when the AI is wrong, but they can also lead to confirmation bias and misinterpretation | AI accuracy | ✓ | ✓ | X | | − |
| (Biessmann & Refiano, 2019) | | image | classification (faces) | classification based on restricted information | human interpretability of SM diverges from automated metrics, easily usable explanations come with the risk of humans replicating AI errors | XAI method, segment size | | | | ✓ | |
| (Cau et al., 2023) | 1 | image | classification (digits) | recognising degraded features | SM do not increase classification accuracy and have mixed effects on agreement with the AI | XAI method, AI accuracy, AI uncertainty, | X | ✓ | ✓ | | + |
| (Chandraseka ran et al., 2018) | 1 | AI | predicting AI accuracy (VQA) | understanding what features influence answers | SM do not increase the accuracy of predicting AI predictions but sometimes decrease it | | X | X | ✓ | X | |
| | 2 | AI | predicting AI prediction (VQA) | understanding what features influence answers | SM do not increase the accuracy of predicting AI predictions but sometimes decrease it | | X | X | ✓ | X | |
| (Chu et al., 2020) | | image | classification (faces) | interpreting features | neither the presence nor the quality of SM affects human accuracy in age prediction | | X | X | ✓ | X | |
| | 1 | AI | predicting AI prediction | bug identification/ bias detection | SM support bias detection, their effects depend on the type and quality of SM, the similarity of discriminative features, and diverge from automated metrics | XAI method, similarity of discriminative features | ✓ | X | ✓ | | |
| (Colin et al., 2022) | 2 | AI | predicting AI prediction | understanding what features enable correct classification | SM support strategy learning in fine-grained classification, their effects depend on the type and quality of SM, the similarity of discriminative features, and diverge from automated metrics | XAI method, similarity of discriminative features | ✓ | X | ✓ | | |
| | 3 | AI | predicting AI prediction | understanding causes of misclassification | SM do not support the understanding of failure cases, their effects depend on the similarity of discriminative features, and diverge from automated metrics | similarity of discriminative features | X | X | ✓ | | |
| (Famiglini et al., 2024) | | image | classification (fractures) | detecting targets | the effects of SM on diagnostic accuracy depend on their visualisation | XAI visualisation, AI accuracy, expertise, task difficulty | | | | ✓ | |
| | 1 | AI | predicting AI prediction | bug identification/ bias detection | SM support bias detection, the combination of where + what is most helpful | XAI method | ✓ | X | X | | – |
| (Fel et al., 2023) | 2 | AI | predicting AI prediction | understanding what features enable correct classification | SM support strategy learning in fine-grained classification, the combination of where + what is most helpful | XAI method | ✓ | X | X | | – |
| | 3 | AI | predicting AI prediction | understanding causes of misclassification | SM do not support the understanding of failure cases, the combination of where + what does not help either | XAI method | X | ✓ | X | | + |
| (Folke et al., 2021) | | AI | predicting AI prediction | understanding what features influence classification | SM support the detection of AI errors, are more helpful for adversarial than normal misclassified images, but are harmful for correctly classified normal images | XAI method, AI accuracy, image type | ✓ | ✓ | X | | + |
| (Giulivi et al., 2021) | | AI | predicting AI prediction | understanding what features influence classification | SM support the prediction of correct AI predictions, the combination of where + what is more helpful for incorrect AI predictions | XAI method, AI accuracy | | | | | 0 |
| (Jin et al., 2024) | | image | classification (glioma) | grading | SM do not increase glioma grading accuracy | | X | X | ✓ | | |
| (John et al., 2021) | | image | classification (faces) | classification based on restricted information | domain-specific SM support image classification more than general ones | XAI method | | | | ✓ | |
| (Jungmann et al., 2023) | | image | classification (breast cancer) | distinguishing targets from distractors | SM only increase diagnostic accuracy when combined with AI support | additional info about AI | X | X | ✓ | | |
| (Khadivpour et al., 2022) | | AI | predicting AI prediction | understanding what features influence classification | SM are less helpful in predicting AI predictions than example-based explanations, types of SM do not differ | XAI method | | | | X | – |
| (Kim et al., 2018) | | AI | reporting reliance on data artefact | bug identification/ bias detection | SM do not support the detection of spurious correlations | | X | ✓ | ✓ | ✓ | – |
| | 1 | AI | agreement with AI prediction | interpreting features | SM make people believe the AI prediction is correct, even when it is not, and induce confirmation bias | XAI method, AI accuracy | ✓ | X | ✓ | | |
| | 2 | AI | agreement with AI prediction | interpreting features | SM make people believe the AI prediction is correct, even when it is not, and induce confirmation bias | XAI method, AI accuracy | ✓ | X | ✓ | | |
| (Kim et al., 2022) | 3 | image, AI | classification (birds), predicting AI prediction | interpreting features, understanding what features influence classification | SM do not support distinguishing between correct and incorrect AI predictions | XAI method, AI accuracy | ✓ | X | ✓ | ✓ | |
| | 4 | image, AI | classification (objects), predicting AI prediction | interpreting features, understanding what features influence classification | SM do not support distinguishing between correct and incorrect AI predictions | XAI method, AI accuracy | ✓ | X | ✓ | ✓ | |
| (Knapič et al., 2021) | | image | classification (bleeding) | distinguishing targets from distractors | SM do not support medical image classification, types of SM do not differ | | X | X | ✓ | X | |
| (Leemann et al., 2023) | | image, AI | classification (food), agreement with AI prediction | interpreting features | SM quality is less important than image ambiguity in calibrating the evaluations of AI predictions | | | | | X | |
| (Lerman et al., 2021) | | AI | predicting object picked by AI, predicting question (VQA) | understanding what features influence answers | standard SM (explanations of individual objects) are less helpful in assessing object relevance than explanations of object relations | XAI method | ✓ | X | ✓ | | – |
| (Li et al., 2021) | 2 | image, AI | classification (objects), agreement with AI prediction | classification based on restricted information | SM effects on classification accuracy depend on SM type | XAI method | | | | | |
| (Lu et al., 2021) | | image | classification (food, animals) | classification based on restricted information | human interpretability of SM diverges from automated metrics | XAI method, segment size, dataset | ✓ | X | X | ✓ | |
| (Mac Aodha et al., 2018) | | image | classification (butterflies, eye diseases, Chinese characters) | understanding what features enable correct classification | SM support category learning in fine-grained classification | | ✓ | X | X | | |
| (Maehigashi et al., 2023a) | | AI | agreement with AI prediction | interpreting features | extended SM increase agreement with AI predictions, regardless of whether they are right or wrong | XAI method, SM quality, task difficulty | X | X | ✓ | ✓ | – |
| (Maehigashi et al., 2023b) | 1 | image | classification (drowsiness, obesity) | interpreting features | the type of classification task determines whether SM quality has positive or negative effects on classification accuracy | XAI quality, classification task | | | | ✓ | |
| (Müller, Thoß, et al., 2024) | 1 | image | classification (scenes) | classification based on restricted information | SM effects on classification accuracy depend on SM type and image type, human interpretability of SM partly diverges from automated metrics | XAI method, image type | | | | ✓ | |
| | 2 | image | classification (scenes) | classification based on restricted information | SM effects on classification accuracy depend on SM type and image type, human interpretability of SM partly diverges from automated metrics | XAI method, image type | | | | ✓ | |
| (Nguyen et al., 2021) | 1 | image | classification (objects) | interpreting features | SM support classification accuracy less than example-based explanations, types of SM do not differ, human interpretability of SM diverges from automated metrics | | X | X | ✓ | X | – |
| | 2 | image | classification (dogs) | interpreting features | SM do not increase classification accuracy but increase the agreement with incorrect AI predictions, human interpretability of SM diverges from automated metrics | | X | ✓ | ✓ | X | – |
| | 3 | image | classification (objects) | interpreting features | also for AI experts, SM support classification accuracy less than example-based explanations | | | | | | – |
| (Park et al., 2018) | | AI | predicting AI accuracy (VQA) | understanding what features influence answers | SM increase the accuracy of predicting AI accuracy | | ✓ | X | X | | |
| (Puri et al., 2019) | | image | chess playing | selecting the best move | SM can improve the selection of chess moves, depending on the specificity and relevance of SM | XAI method | ✓ | ✓ | X | ✓ | |

| | | | | | | | | | | |
|---|---|---|---|---|---|---|---|---|---|---|
| (Qi et al., 2021) | 1 | image | image clustering (birds, cancer cells) | interpreting features | simple SM do not support the clustering of difficult images like the AI, multiple feature-specific SM do | XAI method | X | X | ✓ | | – |
| | 2 | image | image clustering (birds, cancer cells) | interpreting features | simple SM do not support the clustering of difficult images like the AI, multiple feature-specific SM do | XAI method | X | X | ✓ | | – |
| (Ray et al., 2021) | | AI | predicting AI accuracy (VQA) | understanding what features influence answers | SM alone do not improve predictions of AI accuracy, SM combined with error maps do, people predict the AI to be correct when the SM points to relevant regions | XAI method, helpfulness, AI accuracy | X | X | ✓ | | – |
| (Ray et al., 2019) | 1 | AI | predicting image picked by AI (VQA) | understanding what features influence answers | SM improve the accuracy of predicting the image only when they are of excellent quality | XAI method, XAI quality | ✓ | X | ✓ | | – |
| | 2 | AI | predicting image picked by AI (VQA) | understanding what features influence answers | SM do not improve the accuracy of predicting the image when they are good, but are harmful when they are bad | XAI method, XAI quality | X | ✓ | ✓ | | – |
| (Ribeiro et al., 2016) | 3 | AI | reporting reliance on data artefact | bug identification/ bias detection | SM support bias detection | | ✓ | X | X | | |
| (Richard et al., 2023) | | image | classification (mines) | distinguishing targets from distractors | SM do not increase classification accuracy but increase solution time | | X | ✓ | ✓ | | |
| (Sayres et al., 2019) | | image | classification (diabetic retinopathy) | grading | SM speed up diagnosis but do not increase diagnostic accuracy, they increase it when the disease is present but decrease it when it is absent, benefits only occur when the AI is correct | presence of problem (disease), AI accuracy | ✓ | ✓ | ✓ | | |
| (Schuessler & Weiß, 2019) | | image | classification (objects) | classification based on restricted information | SM segments are classified as accurately as human-generated segments | | | | | | 0 |
| (Selvaraju et al., 2017) | | image | classification (objects) | classification based on restricted information | only certain types of SM support classification | XAI method | | | | ✓ | |
| (Shen & Huang, 2020) | 1 | AI | predicting AI prediction | understanding causes of misclassification | SM decrease the accuracy of inferring the misclassified label | misclassificati on type | X | ✓ | ✓ | | |
| | 2 | AI | predicting AI prediction | understanding causes of misclassification | SM decrease the accuracy of inferring the misclassified label | misclassificati on type | X | ✓ | ✓ | | |
| (Shitole et al., 2021) | | AI | predicting relevant areas | understanding what features influence classification | simple SM lead to less accurate but faster selection of relevant regions than structured sets of SM | XAI method | | | | | – |
| (Slack et al., 2021) | 2 | image | classification (digits) | classification based on restricted information | SM provide more informative areas when uncertainty is added | XAI method | | | | ✓ | |
| (Sokol & Flach, 2020) | | AI | predicting relevant areas | understanding what features influence classification | simple SM are less helpful in assessing the influence of features on AI predictions than tree-structured SM | XAI method | ✓ | X | X | | − |
| (Stock & Cisse, 2018) | 2 | AI | agreement with AI prediction | classification based on restricted information | SM increase agreement with AI predictions for misclassified adversarial examples | | X | ✓ | ✓ | | |
| (Sun et al., 2023) | 1 | AI | reporting of challenges of XAI usage | bug identification/ bias detection | SM support bias detection, but working with them is challenging (high effort, uncertainty about meaning, implications for action) | | ✓ | ✓ | X | | |
| | 2 | AI | performance of work strategies | bug identification/ bias detection | SM support the development of more robust AI by mitigating problems of contextual bias | | ✓ | X | ✓ | | |
| (Wang & Vasconcelos, 2023) | 1 | image | classification (birds) | classification based on restricted information | modified SM expose AI insecurities better than randomly cropped regions | | | | | | |
| | 2 | image | classification (birds) | understanding what features enable correct classification | modified SM support category learning in fine-grained classification | | ✓ | X | X | | |
| (Yang et al., 2022) | 1 | AI | predicting AI prediction | understanding what features influence classification | SM increase accuracy in predicting incorrect AI predictions, especially for incorrect AI predictions | AI accuracy | ✓ | X | ✓ | | |
| (Yang et al., 2021) | | AI | predicting AI prediction | understanding what features influence classification | SM help distinguishing between own knowledge and AI knowledge, they support the detection of incorrect AI predictions but are harmful for correct AI predictions | AI accuracy, familiarity | ✓ | ✓ | X | X | |
| (Zhang et al., 2022) | 2 | image | classification (objects) | recognising degraded features | debiasing of SM restores their benefits for images with systematic error | XAI method, blur level | | | | ✓ | |
| (Zhao & Chan, 2023) | 2 | AI | predicting AI prediction (object detected) | understanding what features influence detection | extended SM increase the accuracy of predicting AI predictions, simple SM are less helpful, higher SM quality increases accuracy | XAI method, XAI quality | | | | ✓ | |

*Note.* SM = saliency map, SM+ = benefits of saliency maps, SM− = costs of saliency maps, SM0 = null effects of saliency maps, SMV = differences between versions of saliency maps, SME = differences between standard saliency maps and extensions or combinations with other XAI methods, ✓ = effects are present, X = effects are absent, + = standard saliency maps are superior, − = standard saliency maps are inferior, 0 = no differences or superior in some aspects but inferior in others.